\documentclass{article}
\usepackage{amsmath}
\usepackage{amssymb}
\usepackage{amsthm}
\usepackage{hyperref}
\usepackage[margin=1.0in]{geometry}
\usepackage{multicol}
\usepackage{graphicx}
\usepackage{float}
\usepackage{mathrsfs}
\usepackage{tikz}
\usepackage{shuffle}

\newtheorem{thm}{Theorem}[section]
\newtheorem{lem}{Lemma}[section]
\newtheorem*{cor}{Corollary}

\newcommand{\I}{\infty}

\newcommand{\bt}{\beta}
\newcommand{\gm}{\gamma}

\newcommand{\ep}{\varepsilon}

\newcommand{\et}{\eta}

\newcommand{\ld}{\lambda}
\newcommand{\sm}{\sigma}

\newcommand{\rh}{\rho}
\newcommand{\om}{\omega}

\newcommand{\Dt}{\Delta}

\newcommand{\Ld}{\Lambda}

\newcommand{\Om}{\Omega}

\newcommand{\Z}{{\mathbb{Z}}}
\newcommand{\R}{{\mathbb{R}}}
\newcommand{\C}{{\mathbb{C}}}
\newcommand{\N}{{\mathbb{Z}}_{> 0}}

\newcommand{\id}{{\mathrm{id}}}

\newcommand{\sgn}{{\mathrm{sgn}}}

\newcommand{\limi}[1]{\lim_{#1 \to \infty}}

\newcommand{\ds}{\displaystyle}
\newcommand{\ul}{\underline}
\newcommand{\mft}{\mathfrak{t}}
\newcommand{\mfs}{\mathfrak{s}}
\newcommand{\rmn}[1]{{\rm{#1}}}

\title{Constellation Ensembles and Interpolation in Ensemble Averages}
\author{Elisha D. Wolff}
\date{\today}
\newcommand{\keywords}{Partition function, Berezin integral, Pfaffian, Hyperpfaffian, Grand canonical ensemble, Confluent Vandermonde, Wronskian}

\begin{document}
	
	\maketitle

	\begin{abstract}We introduce constellation ensembles, in which charged particles on a line (or circle) are linked with charged particles on parallel lines (or concentric circles). We present formulas for the partition functions of these ensembles in terms of either the Hyperpfaffian or the Berezin integral of an appropriate alternating tensor. Adjusting the distances between these lines (or circles) gives an interpolation between a pair of limiting ensembles, such as one-dimensional $\bt$-ensembles with $\bt=K$ and $\bt=K^2$.
	\end{abstract}

	
	\section{Introduction}
	\label{sec:intro}
	Suppose a finite number of charged particles are placed on an infinite wire represented by the real line. The charges of the particles are assumed to be integers with the same sign, and the particles are assumed to repel each other with logarithmic interactions. We assumed any two particles of the same charge are indistinguishable. The wire is imbued with a potential which discourages the particles from escaping to infinity in either direction, and heat is applied to the system according to a parameter we call inverse temperature $\bt$. 
	
	Next, suppose this system is copied onto a parallel line (translated vertically in the complex plane). In addition to the internal interactions between particles on the same line, particles from different lines are also able to interact with each other, with the strength of this interaction depending on the distance between the lines. This is an example of what we will call a Linear Constellation Ensemble. We will consider several variations on this setup:
	\begin{enumerate}
		\item \emph{The ($K$-fold) First Constellation Ensemble}, in which charge $L=1$ particles are copied onto $K$ many parallel lines, subject to $\bt=1$.
		\item \emph{The ($K$-fold) Monocharge Constellation Ensemble}, in which particles of the same integer charge $L$ are copied onto $K$ many lines.
		\item \emph{The ($K$-fold) Homogeneous Constellation Ensemble}, in which particles on the same line have the same integer charge $L_k$, but particles on different lines may have different charges.
		\item \emph{The ($K$-fold) Multicomponent Constellation Ensemble}, in which the original line may have particles of different charges, but all the parallel lines are copies, featuring the same charges in the same positions.
	\end{enumerate}
	The first is a special case of the second, which is a special case of either the third or the fourth. Rather than start with the case which is most general (and therefore convoluted), we will work our way up through the different levels of complexity, introducing various tools along the way only as necessary. For each of these ensembles, we will also consider Circular Constellation Ensembles of concentric circles in the complex plane. 
	
	\begin{figure}[h]\label{fig:monocharge}
		\centering
		\caption{A Monocharge (Linear) Constellation Ensemble. }
			\reflectbox{
				\includegraphics[width=0.9\textwidth]{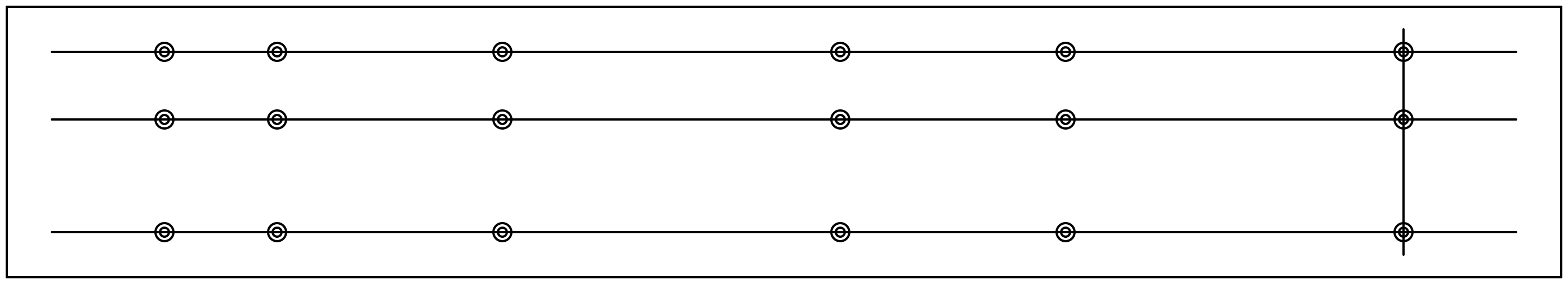}}
	\end{figure}
	
	In Figure 1, there are $K=3$ parallel lines (not necessarily equidistant) on which charge $L=2$ particles have been placed, represented in this figure by pairs of concentric circles. Note, each horizontal line is a copy of the others, so they have the same number of particles at the same (horizontal) locations. Particles which land on the same vertical line are called a \emph{constellation}. In this example, each constellation is made up of $K=3$ particles of the same charge $L=2$. In general, Constellation Ensembles are ensembles of constellations, of which there are $M=6$ in this configuration. 
	
	\begin{figure}[h]\label{fig:homogeneous}
		\centering
		\caption{A Homogeneous (Linear) Constellation Ensemble.  }
		\reflectbox{
			\includegraphics[width=0.9\textwidth]{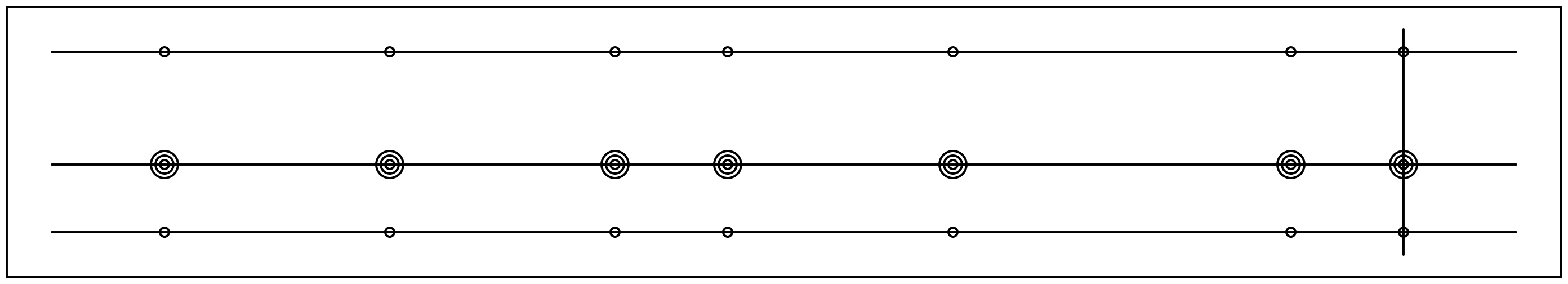}}
	\end{figure}
	
	In Figure 2, there are still $K=3$ parallel lines, but now there are both charge $L_1,L_3=1$ particles and charge $L_2=3$ particles. Note, the top line features only particles of charge $L_3=1$, while the middle line features only particles of charge $L_2=3$. Each constellation (of which there are $M=7$) is made up of one particle of charge 3 and two particles of charge 1, for a total charge of $R_1=5$.
	
	\begin{figure}[h]
		\centering
		\caption{A Multicomponent (Linear) Constellation Ensemble.}
		\reflectbox{
			\includegraphics[width=0.9\textwidth]{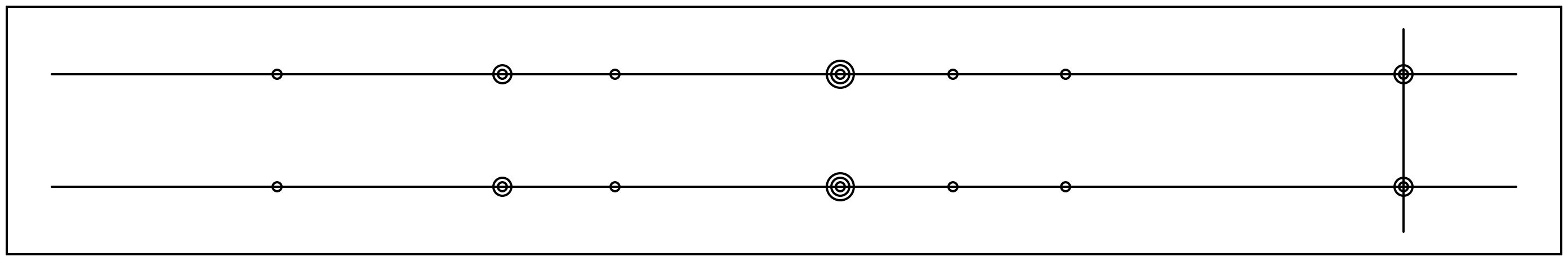}}
	\end{figure}
	
	In Figure 3, each horizontal line features a mix of charge 1, charge 2, and charge 3 particles. However, particles which land on the same vertical line have the same charge. On the left, we've marked a constellation of charge 2 particles. This example is a \emph{multicomponent} ensemble because it is made up of different \emph{species} of constellations, namely $M_1=4$ constellations of charge 1 particles, $M_2=2$ constellations of charge 2 particles, and $M_3=1$ constellation of charge 3 particles. 
	
	\begin{figure}[h]
		\centering
		\caption{A Homogeneous Circular Constellation Ensemble.}
		\includegraphics[width=0.9\textwidth]{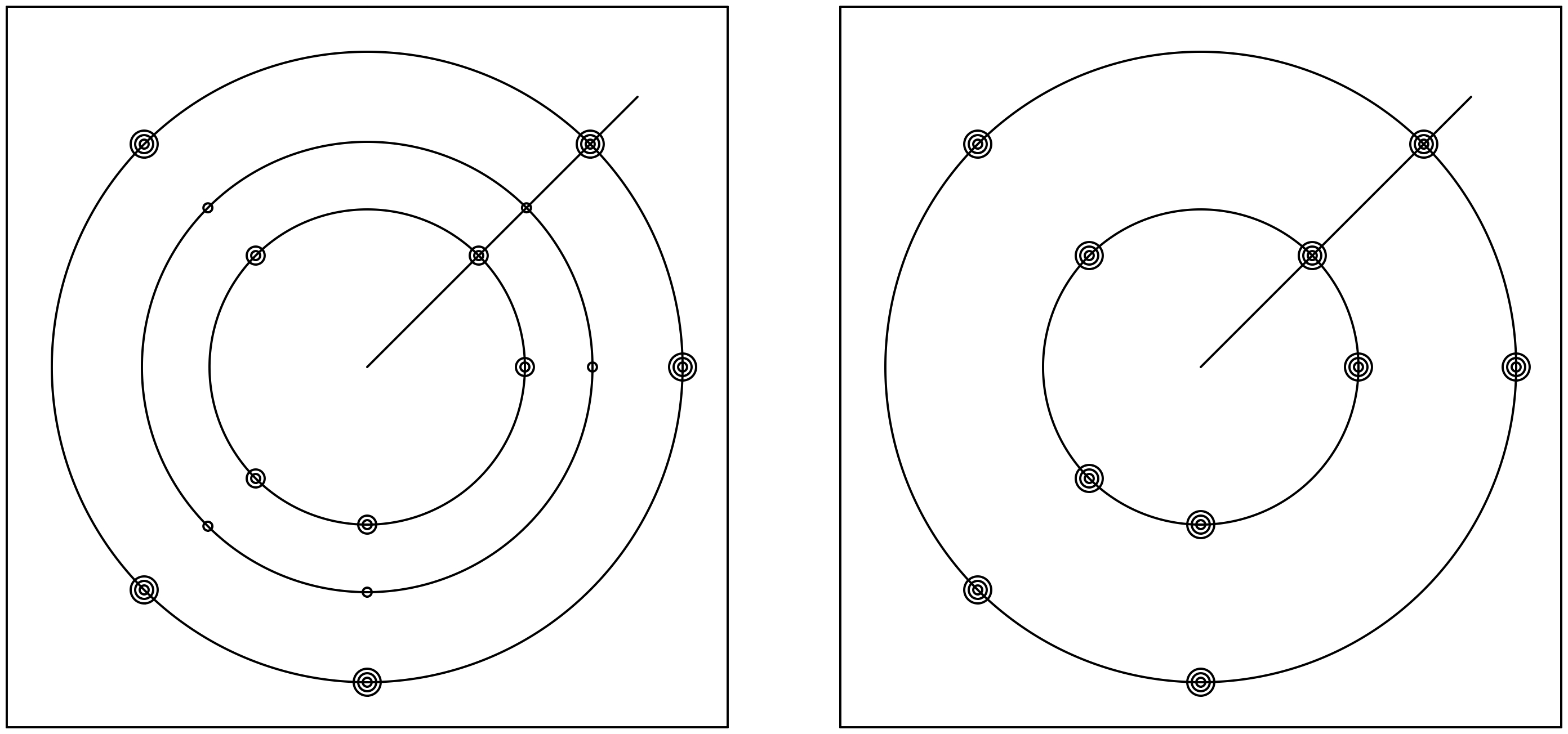}
	\end{figure}
	
	On the left side of Figure 4, there are $K=3$ concentric circles. Note, each constellation (of which there are $M=5$) is made up of particles on the same ray. One such constellation (of three particles) has been marked. The box on the right depicts the result of reducing the radius of the second circle to the radius of the innermost circle. Each charge 1 particle merges with a charge 2 particle to form a charge $1+2=3$ particle. 
	
	Though these particle arrangements are somewhat contrived physically, the resulting joint probability density functions give us insight into limiting ensembles which we can interpolate between (by adjusting the distances between the parallel lines or circles). For example, taking the limit of the First Constellation Ensemble as the distance between the lines (or circles) goes to zero (and correcting for the singularities as particles collapse onto each other) produces a one-dimensional $\bt=K^2$ ensemble. On the other end, taking the limit as the distance between the lines (or circles) goes to infinity produces a one-dimensional $\bt=K$ ensemble.
	
	Previously, in \cite{wolff2021partition}, we gave generalizations (included in \autoref{sec:gendb}) of the de Bruijn integral identities \cite{deBruijn1955}, in which the iterated integral of a determinant is expressed as the Hyperpfaffian or Berezin integral (see \autoref{ssec:berezin}) of an appropriate alternating tensor (also \emph{form}). As the first application, we substitute the particulars for the partition function of the Monocharge Constellation Ensemble in \autoref{sec:monocharge}. In \autoref{sec:homogen}, we extend this to Homogeneous Constellation Ensembles, the most general classification (in this volume) which still produces homogeneous forms (and therefore Hyperpfaffian partition functions). Conversely, in \autoref{sec:multi}, we consider Multicomponent Constellation Ensembles which produce non-homogeneous forms instead. Finally, in \autoref{sec:circ}, we consider (circular) ensembles of concentric circles in place of parallel lines. In all cases, the generalized de Bruijn identities are used, further demonstrating the versatility in the methods established in our previous volume. 
	
	\subsection{Historical context}
	\label{sec:hist}
	
	
	The $\bt$-ensembles are a well-studied collection of random matrices whose eigenvalue densities take a common form, indexed by a non-negative, real parameter $\bt$. First, let $\Dt(\vec{x})$ denote the Vandermonde determinant in variables $x_1,\ldots,x_N$, so that
	\[
	\Dt(\vec{x})=\prod_{i<j}(x_j-x_i).
	\] 
	Next, suppose $\mu$ is a continuous probability measure on $\R$ with Radon-Nikodym derivative $\frac{d \mu}{d x} = w(x)$. For each $\bt\in \R_{> 0}$, consider the $N$-point process specified by the joint probability density
	\[
	\rh_N(x_1,\ldots,x_N)  =\frac{1}{Z_{N}(\bt)N!}\left|\Dt(\vec{x})\right|^\bt \prod_i w(x_i)
	\]
	where $Z_N(\bt)$, which denotes the \emph{partition function} of $\bt$, is the normalizing constant required for $\rh_N$ to be a probability density function. This eigenvalue density function can be identified with the Boltzmann factor of the previously discussed log-gas particles, as first observed by Dyson \cite{Dyson1962}, and further developed by Forrester in \cite{Forrester2010}. 
	
	The \emph{classical} $\bt$-ensembles (with $\bt=1,2,4$ and $w(x) = e^{-x^2/2}$), corresponding to Hermitian matrices with real, complex, or quaternionic Gaussian entries (respectively), were first studied in the 1920s by Wishart in multivariate statistics \cite{wish} and the 1950s by Wigner in nuclear physics \cite{wig}. In the subsequent decade, Dyson and Mehta \cite{Dyson1963} unified a previously disparate collection of random matrix models by demonstrating that the three classic $\bt$-ensembles are each variations of a single action on random Hermitian matrices (representing the three associative division algebras over $\R$). In \cite{Dumitriu2002}, Dumitriu and Edelman provide tridiagonal matrix models for $\bt$-ensembles of arbitrary positive $\bt$, which are then used by Ram\'irez, Rider, and Vir\'ag in \cite{Ramirez2011} to obtain the asymptotic distribution of the largest eigenvalue.   
	
	For each $1\leq n\leq N$, define the $n^{\rm{th}}$ correlation function by 
	\[
	R_n(x_1,\ldots,x_n)=\frac{1}{(N-n)!}\int_{\R^{N-n}}\rh_N(x_1,\ldots,x_n,y_1,\ldots,y_{N-n})\,dy_1\cdots dy_{N-n}.
	\]
	It turns out that the correlation function for the classic $\beta$-ensembles takes a particularly nice algebraic form. For example, when $\bt=2$, it can be shown using only elementary matrix operations and Fubini's Theorem that
	\[
	R_n(x_1,\ldots,x_n)=\frac{1}{Z_N(2)}\det(K(x_i,x_j)_{1\leq i,j\leq n}),
	\]
	where the \emph{kernel} $K(x,y)$ is a certain square integrable function $\R\times \R\to \R$ that can most easily be expressed in terms a family of polynomials which are orthogonal with respect to the measure $\mu$. For this reason, we say the classical $\bt=2$ ensemble is an example of a \emph{determinantal} point process. The details of this derivation are given in \cite{Mehta2004}. Similarly, when $\bt=1$ or $4$, 
	\[
	R_n(x_1,\ldots,x_n)=\frac{1}{Z_N(\bt)}{\rm{Pf}}(K(x_i,x_j)_{1\leq i,j\leq n}),
	\]
	where ${\rm{Pf}}(A)=\sqrt{\det(A)}$ denotes the Pfaffian of an antisymmetric matrix $A$, and where $K(x,y)$ is a certain $2\times 2$ matrix-valued function whose entries are square-integrable, and which satisfies $K(x,y)^T = -K(y,x)$. We then say the classical $\bt=1$ and $\bt=4$ ensembles are examples of \emph{Pfaffian} point processes. This result was first shown for circular ensembles by Dyson in \cite{Dyson1962}, then for Gaussian ensembles by Mehta in \cite{Mehta2004} and then for general weights ($\mu$) by Mehta and Mahoux in \cite{mahoux}, except for the case $\bt=1$ and $N$ odd. Finally, the last remaining case was given by Adler, Forrester, and Nagao in \cite{adler}. Of fundamental concern in the theory of random matrices is the behavior of eigenvalue statistics as $N\to \I$. The immediate advantage of these determinantal and Pfaffian expressions for the correlation functions is that these matrix kernels do not essentially increase in complexity as $N$ grows large, since the dimensions are the matrix kernel are stable, and the entries are expressed as a sum whose asymptotics are well-understood.
	

	\subsection{Hyperpfaffian partition functions}
	\label{sec:hpf}
	
	Derivations of the determinantal and Pfaffian expressions of the correlation functions have been presented in numerous ways over the past several decades. Of particular note is the method of Tracy and Widom \cite{wid}, who first show that the partition function is determinantal or Pfaffian, and then use matrix identities and generating functions to obtain a corresponding form for the correlation functions.	
	
	But recognizing the partition function $Z_N(\bt)$ as the determinant or Pfaffian of a matrix of integrals of appropriately chosen orthogonal polynomials is essential and nontrivial. One way to do this is to apply the Andreif determinant identity \cite{Andreief1884} to the iterated integral which defines $Z_N(\bt)$. This is immediate when $\bt=2$, and viewing the Pfaffian as the square root of a determinant, this identity can also be applied (with some additional finesse) when $\bt=1$ or $4$. However, viewing the Pfaffian in the context of the exterior algebra allows us to extend the Andreif determinant identity to analogous Pfaffian identities (referred to as the de Bruijn integral identities). 
	
	In 2002, Luque and Thibon \cite{Luque2002} used techniques in the shuffle algebra to show that when $\bt=L^2$ is an even square integer, the partition function $Z_N(\bt)$ can be written as a Hyperpfaffian of an $L$-form whose coefficients are integrals of Wronskians of suitable polynomials. Then in 2011, Sinclair \cite{Sinclair2011} used other combinatorial methods to show that the result also holds when $\bt=L^2$ is an odd square integer.

	In his 2013 dissertation, Shum \cite{Shum2013solvable} considered 2-fold First Cosntellation Ensembles (both linear and circular), demonstrating these ensembles to be completely solvable Pfaffian point processes. Additionally, he showed how these ensembles give an interpolation between the classical $\bt=2$ and $\bt=4$ ensembles. 
	In this volume, the many new variations on the constellation setup allow for many more interpolations, including but not limited to an interpolation between $\bt=L$ and $\bt=L^2$ ensembles. Thus, the partition functions of integer $\bt$-ensembles can all be written as a limit of Hyperpfaffians, even when $\bt$ is a square-free integer. 
	

	\subsection{The monocharge setup}
	\label{ssec:setup}
	Let $\vec{x}\in \R^M$, and let $\vec{y}\in \R^K$ such that $0\leq y_1<\cdots <y_K$. We call $\vec{y}$ the \emph{translation vector} of the system, giving the locations of the $K$ many lines $\R+iy_k$ in the complex plane. Consider $M$ many charge $L\in \Z_{>0}$ particles on each line $\R+iy_k$ having the same real parts, meaning for each location $x_m\in \R$, and $1\leq k\leq K$, there is a charge $L$ particle at location $x_m+iy_k$. Denote the (total $KM$) particle locations by
	\[
	{\textbf{x}}=(\textbf{x}^1,\textbf{x}^2,\ldots,\textbf{x}^M)\in \C^{KM},
	\]
	where $\textbf{x}^m=x_m+i\vec{y}=(x_m+iy_1,x_m+iy_2,\ldots,x_m+iy_K)\in \C^K$. We call ${\textbf{x}}$ the \emph{location vector} of the system, in which each $x_k^m\in \C$ gives the location of a particle. We call $\textbf{x}^m$ the location vector of the \emph{constellation} of $K$ many particles which all share the same real part $x_m$. We call $\vec{x}=(x_1,\ldots,x_M)$ the location vector of the real parts which generate each constellation. 
	
	The particles are assumed to interact logarithmically so that the contribution of energy to the system by two (charge $L$) particles at locations $x_m+iy_k$ and $x_n+iy_j$ is given by $-L^2\log|(x_m+iy_k)-(x_n+iy_j)|$. Let $U:\R\to \R$ be a potential on the real axis. Let $\overline{U}:\C\to\R$ be a extension of this potential to the entire complex plane such that $\overline{U}(z)=U(\rmn{Re}(z))$. Without loss of generality, we can assume $x_1<\ldots<x_M$. Then at inverse temperature $\bt$, the total potential energy of the system is given by
	\begin{align*}
		E(\vec{x},\vec{y})&=\bt L\sum_{k=1}^K\sum_{m=1}^M\overline{U}(x_m+iy_k)-\bt L^2\sum_{k=1}^K\sum_{n<m}^{M}\log|(x_m+iy_k)-(x_n+iy_k)|\\&\hspace{5mm}-\bt L^2\sum_{j<k}^{K}\sum_{m=1}^{M}\log|(x_m+iy_k)-(x_m+iy_j)|\\&\hspace{5mm}-\bt L^2\sum_{j<k}^K\sum_{n<m}^M\log|(x_m+iy_k)-(x_n+iy_j)|+\log|(x_m+iy_j)-(x_n+iy_k)|.
	\end{align*}
	The first iterated sum in the first line accounts for the potential $\overline{U}$. We can substitute $\overline{U}(x_m+iy_k)=U(x_m)$ of which there are $K$ many for each $m$. The second iterated sum in the first line accounts for interactions between particles which share a line. Note, the differences in that iterated sum are all positive by assumption on the ordering of the $x_m$, and the differences are the same for all $1\leq k\leq K$. The iterated sum in the second line accounts for interactions between particles of the same constellation, meaning same real part $x_m$. The differences in that iterated sum are the same for $1\leq m\leq M$. The iterated sum in the third line accounts for the remaining interactions between particles. For each quadruple $(m,k,n,j)$, we get four points which make up a rectangle in the complex plane. The four sides of this rectangle are already accounted for by the other interactions. The product of the lengths of the two diagonals is the sum of the squares of the lengths of the sides. Thus, the potential energy simplifies to
	\begin{align*}
		E(\vec{x},\vec{y})&=\bt L K\sum_{m=1}^MU(x_m)-\bt L^2K\sum_{n<m}^{M}\log(x_m-x_n)-\bt L^2 M\sum_{j<k}^K\log |i(y_k-y_j)|\\&\hspace{5mm}-\bt L^2\sum_{j<k}^K\sum_{n<m}^M\log\left((x_m-x_n)^2+(y_k-y_j)^2\right).
	\end{align*}
	
	With this setup, the relative density of states (corresponding to varying location vectors $\vec{x}$ and translation vectors $\vec{y}$) is given by the Boltzmann factor
	\[
	\Om(\vec{x},\vec{y})=\exp(-E(\vec{x},\vec{y}))=|\Dt({\textbf{x}})|^{\bt L^2}\prod_{m=1}^Me^{-\bt LKU(x_m)}=\Dt({\textbf{x}})^{\bt L^2}\prod_{m=1}^M\left((-i)^{L(K-1)/2}e^{-U(x_m)}\right)^{\bt LK},
	\]
	where $\Dt({\textbf{x}})$ denotes the Vandermonde determinant, evaluated at the variables ${\textbf{x}}$. Note, the last equality comes from $|i|=(i)(-i)$, of which there are $\bt L^2M{K \choose 2}$ many instances. Thus, the probability of finding the system in a state corresponding to a location vector $\vec{x}$ and fixed translation vector $\vec{y}$ is given by the joint probability density function
	\[
	\rh(\vec{x},\vec{y})=\frac{\Om(\vec{x},\vec{y})}{Z_M(\vec{y})},
	\]
	where the \emph{partition function} (of the $K$-fold Monocharge Constellation Ensemble) $Z_M(\vec{y})$ is the normalization constant given by
	\begin{align*}
		Z_{M}(\vec{y})&=\int_{-\I<x_1<\ldots<x_M<\I}\Om(\vec{x},\vec{y})\,dx_1\cdots dx_M\\&=\int_{-\I<x_1<\ldots<x_M<\I}\Dt({\textbf{x}})^{\bt L^2}\,d\mu(x_1)\cdots d\mu(x_M),
	\end{align*}
	in which $d\mu(x)=\left((-i)^{L(K-1)/2}e^{-U(x)}\right)^{\bt LK}dx$. At this point, it is necessary to assume the potential $U$ is one for which $Z_M(\vec{y})$ is finite. 
	
	Unit charges (meaning $L=1$) at inverse temperature $\bt=b^2$ have the same Boltzmann factor (and resulting density function) as charge $L=b$ particles at inverse temperature $\bt=1$ (subject to different but related potentials $U(x)$). In general, replacing $\bt$ with $\bt'=\bt/b^2$ and replacing $L$ with $L'=bL$ leaves $\Dt({\textbf{x}})^{\bt L^2}$ unchanged. Then replacing $U$ with $U'=bU$ leaves $\Om(\vec{x},\vec{y})$ unchanged. Thus, for computational purposes, we can change to $\bt=1$ (provided $\sqrt{\bt}L\in \Z$ for the original $\bt$). 
	
	The partition function $Z_{M}(\vec{y})$ and its analogues are the primary objects of interest to us. Though we assume $\bt=1$ for computational purposes, $Z_{{M}}(\vec{y})$ is inherently a function of $\bt$, among other parameters. The potential $U$ dictates the external forces experienced by each particle individually, affecting the (complex) measures $\mu$ against which we are integrating. The charge $L$ and the inverse temperature $\bt$ influence the strength of the interactions between the particles, affecting the exponents on the interaction terms in the Boltzmann factor. As $y_k-y_j\to 0$, the corresponding interaction terms shrink, and the potential energy grows. Conversely, as $y_k-y_j\to \I$, the corresponding interaction terms grow, and the potential energy shrinks. 
	
	Recall, this $Z_M(\vec{y})$ is an iterated integral in $M$ many variables. As in the previous volume (Wolff and Wells 2021), our goal here is not to compute these integrals for any particular choice of several parameters. Instead, we demonstrate, in general, how to write $Z_M(\vec{y})$ as a Hyperpfaffian (or Berezin integral in the multicomponent case) of a form whose coefficients are only single or double integrals of (potentially orthogonal) polynomials.

	\section{Preliminary definitions}
	\label{sec:prelim}
	In this section, we introduce a mix of conventions and definitions which simplify the statement of our main results. First, for any positive integer $N$, let $\underline{N}$ denote the set $\{1,\ldots,N\}$. Assuming positive integers $K\leq N$, let $\mft:\underline{K}\nearrow\underline{N}$ denote a strictly increasing function from $\underline{K}$ to $\underline{N}$, meaning
	\[
	1\leq\mft(1)<\mft(2)<\cdots<\mft(K)\leq N.
	\]
	It will be convenient to use these increasing functions to track indices used in denoting minors of matrices and elements of exterior algebras, among other things (often in place of, but sometimes in conjunction with, permutations). For example, given an $N\times N$ matrix $V$, $V_{\mft}$ might denote the $K\times K$ minor composed of the rows $\mft(1),\ldots,\mft(K)$, taken from the first $K$ columns of $V$.

	\subsection{Wronskians}
	\label{ssec:wr}
	For any non-negative integer $l$, define the $l^{\rmn{th}}$ modified differential operator $D^l$ by
	\[
	D^lf(x)=\frac{1}{l!}\frac{d^lf}{dx^l},
	\]
	with $D^0f(x)=f(x)$. Define the modified Wronskian, Wr$(\vec{f},x)$, of a family, $\vec{f}=\{f_n\}_{n=1}^L$, of $L$ many sufficiently differentiable functions by 
	\[
	\rmn{Wr}(\vec{f},x)=\det\left[D^{l-1}f_n(x)\right]_{n,l=1}^L.
	\]
	We call this the \emph{modified} Wronskian because it differs from the typical Wronskian (used in the study of elementary differential equations to test for linear dependence of solutions) by a combinatorial factor of $\prod_{l=1}^L l!$. 
	
	A \emph{complete} $N$-family of monic polynomials is a collection $\vec{p}=\{p_n\}_{n=1}^N$ such that each $p_n$ is monic of degree $n-1$. Given $\mft:\underline{L}\nearrow \underline{N}$, define $\vec{p}_\mft=\{p_{\mft(j)}\}_{j=1}^L$. Then the (modified) Wronskian of $\vec{p}_\mft$ is given by
	\[
	\rmn{Wr}(\vec{p}_\mft,x)=\det\left[D^{l-1}p_{\mft(j)}(x)\right]_{j,l=1}^L.
	\]
	Similarly, define the proto-Wronskian, $\rmn{Pr}_{\vec{y}}(\vec{f},x)$, (with respect to translation vector $\vec{y}$) by
	\[
	\rmn{Pr}_{\vec{y}}(\vec{f},x)=\det\left[f_n(x+iy_k)\right]_{n,k=1}^K.
	\]
	We call this the \emph{proto}-Wronskian because 
	\[
	\lim_{\vec{y}\to 0} \frac{\rmn{Pr}_{\vec{y}}(\vec{f},x)}{\Dt (i\vec{y})}=\rmn{Wr}(\vec{f},x).
	\]
	A proof of this is given in \autoref{ssec:proto}. The Wronskian, which appears when studying one-dimensional ensembles, has columns generated by taking higher derivatives of each $f_n$. The number of columns is equal to the charge of the particles under consideration. The proto-Wronskian, which appears when studying First Linear Constellation Ensembles, has columns generated by instead evaluating each $f_n$ at different translations $x+iy_k$. The number of columns $K$ is equal to the number of parallel lines under consideration. 
	
	In the case of the Monocharge Constellation Ensemble $(L\neq 1)$, it is necessary to conflate these two structures. To that end, for $\vec{f}=\{f_m\}_{m=1}^{LK}$, define 
	\[
	\rmn{Wr}\otimes \rmn{Pr}_{\vec{y}}(\vec{f},x)=\det \left[\left[D^{l-1}f_{(n-1)L+j}(x+iy_k)\right]_{j,l=1}^L\right]_{n,k=1}^K.
	\]
	The first column of the associated matrix is $LK$ many functions evaluated at $x+iy_1$. The second column is the first derivatives of those functions evaluated at the same $x+iy_1$, and so on until the first $L$ many columns have been exhausted. The next $L$ many columns are the same functions and derivatives evaluated at $x+iy_2$, and so on until all $y_k$ have been exhausted. The resulting $LK\times LK$ matrix will have $L\times L$ Wronskian blocks evaluated at one of the $K$ many $x+iy_k$. In \autoref{ssec:proto}, we will show
	\[
	\lim_{\vec{y}\to 0}\frac{\rmn{Wr}\otimes \rmn{Pr}_{\vec{y}}(\vec{f},x)}{\Dt (i\vec{y})^{L^2}}=\rmn{Wr}(\vec{f},x).
	\]
	Suppose, for example, $L=3$, $K=2$, and $\vec{f}=\{x^{n-1}\}_{n=1}^6$ (which happens when there are 2 parallel lines of charge 3 particles). Then 
	\[
	\rmn{Wr}\otimes \rmn{Pr}_{\vec{y}}(\vec{f},x)=\begin{bmatrix}
		1 & 0 & 0 & 1 & 0 & 0 \\
		x+iy_1 & 1 & 0 & x+iy_2 & 1 & 0 \\
		(x+iy_1)^2 & 2(x+iy_1) & 1 & (x+iy_2)^2 & 2(x+iy_2) & 1 \\
		(x+iy_1)^3 & 3(x+iy_1)^2 & 3(x+iy_1) & (x+iy_2)^3 & 3(x+iy_2)^2 & 3(x+iy_2) \\
		(x+iy_1)^4 & 4(x+iy_1)^3 & 6(x+iy_1)^2 & (x+iy_2)^4 & 4(x+iy_2)^3 & 6(x+iy_2)^2 \\
		(x+iy_1)^5 & 5(x+iy_1)^4 & 10(x+iy_1)^3 & (x+iy_2)^5 & 5(x+iy_2)^4 & 10(x+iy_2)^3 
	\end{bmatrix}.
	\]

	\subsection{The Berezin integral}
	\label{ssec:berezin}
	Let $\ep_1,\ldots,\ep_N$ be a basis for $\R^N$. For any injection $\mft:\underline{K}\to \underline{N}$, let $\ep_\mft\in \bigwedge^{K}(\R^N)$ denote
	\[
	\ep_\mft=\ep_{\mft(1)}\wedge\ep_{\mft(2)}\wedge\cdots\wedge\ep_{\mft(K)}.
	\]
	Then $\{\ep_\mft\,|\,\mft:\underline{K}\nearrow\underline{N}\}$ is a basis for $\bigwedge^K(\R^N)$. In particular, $\bigwedge^N(\R^N)$ is a one-dimensional subspace we call the \emph{determinantal line}, spanned by
	\[
	\ep_{\rmn{vol}}=\ep_{\rmn{id}}=\ep_1\wedge\ep_2\wedge\cdots\wedge\ep_N,
	\]
	which we call the \emph{volume form} (in $\R^N$). For each $0<n\leq N$, define $\frac{\partial}{\partial \ep_n}:\bigwedge^K(\R^N)\to \bigwedge^{K-1}(\R^N)$ on basis elements by 
	\[
	\frac{\partial}{\partial \ep_n}\ep_{\mft}=\begin{cases}
		(-1)^{k}\ep_{\mft(1)}\wedge\cdots\wedge\ep_{\mft(k-1)}\wedge\ep_{\mft(k+1)}\wedge\cdots\wedge\ep_{\mft(K)} & \text{ if } k=\mft^{-1}(n) \\ 0 & \text{ otherwise}
	\end{cases},
	\]
	and then extend linearly. If $n\in \mft(\underline{K})$, meaning $\ep_n$ appears as a factor in $\ep_{\mft}$, then $\frac{\partial \ep_{\mft}}{\partial \ep_n}$ is the result of permuting $\ep_n$ to the front and then removing it, picking up a sign associated with changing the order in which the basis elements occur. If $\ep_\mft$ does not have $\ep_n$ as a factor, then $\frac{\partial \ep_{\mft}}{\partial \ep_n}=0$. Given an injection $\mathfrak{s}:\underline{L}\to \underline{N}$, we define the Berezin integral \cite{Berezin1966} (with respect to $\ep_{\mathfrak{s}}$) as a linear operator $\bigwedge(\R^N)\to \bigwedge(\R^N)$ given by
	\[
	\int\ep_{\mft}\,d\ep_{\mathfrak{s}}=\int \ep_{\mft}\,d\ep_{\mathfrak{s}(1)}\,d\ep_{\mathfrak{s}(2)}\cdots d\ep_{\mathfrak{s}(L)}=\frac{\partial}{\partial \ep_{\mathfrak{s}(L)}}\cdots \frac{\partial}{\partial \ep_{\mathfrak{s}(2)}}\frac{\partial}{\partial \ep_{\mathfrak{s}(1)}}\ep_{\mft}.
	\]
	Our main results are stated in terms of Berezin integrals with respect to the volume form $\ep_{\rmn{vol}}\in \bigwedge^N(\R^N)$. Note, if $\ep_{\mft}\in \bigwedge^K(\R^N)$ for any $K<N$, then 
	\[
	\int \ep_{\mft}\,d\ep_{\rmn{vol}}=0
	\]
	because $\ep_{\mft}$ is missing some $\ep_k$ as a factor. Thus, the Berezin integral with respect to $\ep_{\rmn{vol}}$ is a projection operator $\bigwedge(\R^N)\to \bigwedge^N(\R^N)\cong \R$. In particular, if $\sm\in S_N$, then
	\[
	\int \ep_{\sm}\,d\ep_{\rmn{vol}}=\sgn (\sm).
	\]

	\subsection{Exponentials of forms}
	\label{ssec:exp}
	For $\om\in \bigwedge(\R^N)$ and positive integer $m$, we write
	\[
	\om^{\wedge m}=\om\wedge\cdots\wedge \om,
	\]
	with $\om$ appearing as a factor $m$ times. By convention, $\om^{0}=1$. We then define the exponential
	\[
	\exp(\om)=\sum_{m=0}^\I\frac{\om^{\wedge m}}{m!}.
	\]
	Moreover, suppose $\om=\om_1+\om_2+\cdots+\om_J$ where each $\om_j\in \bigwedge^{L_j}(\R^N)$ and each $L_j$ even, then (we say each $\om_j$ is a homogeneous even form of length $L_j$ and) it is easily verified 
	\[
	\exp({\om})=\exp({\om_1+\cdots+\om_J})=\exp({\om_1})\wedge\cdots\wedge \exp({\om_J}).
	\]
	We get a homogeneous form in all cases but the Multicomponent Constellation Ensemble. In the homogeneous cases, exactly one summand in the exponential will live at the determinantal line. Assuming $\om\in \bigwedge^K(\R^N)$ with $KM=N$, we get
	\[
	\int \exp(\om)\,d\ep_{\rmn{vol}}=\int \sum_{m=0}^\I\frac{\om^{\wedge m}}{m!}\,d\ep_{\rmn{vol}}=\int \frac{\om^{\wedge M}}{M!}\,d\ep_{\rmn{vol}}=\rmn{PF}(\om),
	\]
	where PF$(\om)$ is the \emph{Hyperpfaffian} of $\om$, the real number coefficient on $\ep_{\rmn{vol}}$ in $\frac{\om^{\wedge M}}{M!}$. Thus, this Berezin integral is the appropriate generalization of the Hyperpfaffian. To avoid confusing this Berezin integral with other integrals which appear in our computations, we will write
	\[
	\rmn{BE}_{\rmn{vol}}(\om)=\int\exp(\om)\,d\ep_{\rmn{vol}},
	\]
	where the subscript on the left hand side indicates which form we are integrating with respect to.
	
	The partition function of a one-dimensional ensemble with a single species has been shown to have a Hyperpfaffian expression (for certain $\bt$) \cite{Sinclair2011}. More generally, we showed (Wolff and Wells 2021) the partition function of a one-dimensional ensemble with multiple species can be expressed as the Berezin integral of an exponential. Using the same methods, we obtain a Berezin integral expression for the partition function of all Constellation Ensembles. In the case of Homogeneous Constellation Ensembles, we additionally get a Hyperpfaffian expression. 
	
	\section{Generalized de Bruijn identities}
	\label{sec:gendb}
	
	Let $N=L_1+\cdots+L_J$. Define $K_j=\sum_{k=1}^{j}L_k$. Let $A(\vec{x})$ be an $N\times N$ matrix whose entries are single variable integrable functions of variables $\vec{x}=(x_1,\ldots,x_J)$. Explicitly, the first $L_1$ many columns are functions of $x_1$, the second $L_2$ many columns are functions of $x_2$, and so on up through $x_J$. For $\mft:\ul{L_j}\nearrow\ul{N}$, let $A_\mft(x_j)$ denote the $L_j\times L_j$ minor of $A(\vec{x})$ given by
	\[
	A_\mft(x_j)=\left[A(\vec{x})_{\mft(l),n+K_{j-1}}\right]_{l,n=1}^{L_j},
	\]
	equivalently obtained from $A(\vec{x})$ by taking the rows $\mft(1),\ldots,\mft(L_j)$ from the $L_j$ many columns in the same variable $x_j$. Define
	\[
	\gm_j^A=\sum_{\mft:\ul{L_j}\nearrow \ul{N}}\int_{\R}\det A_\mft (x_j)\,dx_j\,\ep_{\mft},
	\]
	and define 
	\[
	\et_{j,k}^A=\sum_{\mft:\ul{L_j}\nearrow \ul{N}}\sum_{\mathfrak{s}:\ul{L_k}\nearrow \ul{N}}\int\int_{x_j<x_k}\det A_\mft (x_j)\cdot \det A_{\mathfrak{s}} (x_k)\,dx_j\,dx_k\,\ep_{\mft}\wedge \ep_{\mathfrak{s}}.
	\]
	Then the relevant results of our previous volume can be summarized in this theorem. 
	\begin{thm}\label{thm:debruijngen} Suppose the first $r$ many $L_j$ are even, then
		\[
		\int_{-\I<x_1<\ldots<x_J<\I}\det A(\vec{x})\,dx_1\cdots dx_J=\int \om\,d\ep_{\rmn{vol}},
		\]
		where $\om$ is defined as follows:
		\begin{enumerate}
			\item If $N$ is even, then
			\[
			\om=\frac{1}{\left(r+\frac{J-r}{2}\right)!}\bigwedge_{j=1}^{r}\gm_j^A\wedge \bigwedge_{m=1}^{(J-r)/2}\et_{r+2m-1,r+2m}^A.
			\]
			\item If $N$ is odd, then
			\[
			\om=\frac{1}{\left(r+1+\frac{J-r-1}{2}\right)!}\bigwedge_{j=1}^{r}\gm_j^A\wedge \bigwedge_{m=1}^{(J-r-1)/2}\et_{r+2m-1,r+2m}^A\wedge \gm_J^A.
			\]
		\end{enumerate}
	\end{thm}
	
	Note, we require even forms (possibly either $\gm_j^A$ or $\et_{j,k}^A$) so they commute. For $1\leq j\leq r$, $L_j$ is even, and $\gm_j^A$ is an even $L_j$-form. For the $L_j$ which are odd, $\et_{j,k}^A$ combines minors of odd $L_j\times L_j$ dimensions with minors of odd $L_k\times L_k$ dimensions to produce an even $(L_j+L_k)$-form. In case 1, the requirement that $N$ be even means there are an even number of odd $L_j$ to be paired down into $(J-r)/2$ pairs. In case 2, there are an odd number of odd $L_j$, so $\gm_J^A$ remains as an odd $L_J$-form. Though this extra $\gm_J^A$ is an odd form, it commutes with all the even forms. 
	
	In our applications, it is necessary to extend the $\ep_j$ basis for $\R^N$ to a basis for $\R^{N+k}$ and extend the odd $\gm^A_J$ form by these new basis vectors to create another even form. In general, we can write
	\[
	\ep_{\rmn{vol}_k}=\ep_{\rmn{vol}}\wedge\xi_k=\ep_{\rmn{vol}}\wedge \ep_{N+1}\wedge \ep_{N+2}\wedge \cdots \wedge \ep_{N+k}.
	\]
	Then for any $\om\in\bigwedge(\R^N)\leq \bigwedge(\R^{N+k})$, we have
	\[
	\int\om\,d\ep_{\rmn{vol}}=\int\om\wedge \ep_{N+1}\wedge\cdots \wedge\ep_{N+k}\,\,d\ep_{\rmn{vol}}\,d\ep_{N+1}\cdots d\ep_{N+k}=\int\om\wedge \xi_{k}\,\,d\ep_{\rmn{vol}_k}.
	\]
	Thus, we can embed any Berezin integral computation in a higher dimension if desired.

	Recall, we assume the functions which make up $A(\vec{x})$ are suitably integrable so that all integrals which appear in $\gm_j^A$ and $\et_{j,k}^A$ are finite. However, we do not assume any resemblance between the $L_j$ many columns in $x_j$ and the $L_k$ many columns in $x_k$. Assuming some additional consistency, we obtain a Hyperpfaffian analogue of the de Bruijn integral identities.
	
	\begin{cor}\label{cor:debruijnpf} Let $\xi_k=\ep_{N+1}\wedge \ep_{N+2}\wedge\cdots\wedge\ep_{N+k}$. Suppose $L_1=\cdots=L_J=L$. Under the additional assumption that $\gm_j^A=\gm$ for all $j$, and $\et_{j,k}^A=\et$ for all $j,k$ (typically because the entries of $A(\vec{x})$ in one variable $x_j$ are the same as the entries in any other variable $x_k$), 
		\[
		\int_{-\I<x_1<\ldots<x_J<\I}\det A(\vec{x})\,dx_1\cdots dx_J={\rm{BE}}_{\rmn{vol}_k}(\om)={\rm{PF}}(\om),
		\]
		where $\om$ and $k$ depend on $M$ and $L$.
		\begin{enumerate}
			\item If $L$ is even, then $\om=\gm$ and ${\rm{BE}}_{\rmn{vol}_k}={\rm{BE}}_{\rmn{vol}}$.
			\item If $L$ is odd and $M$ is even, then $\om=\et$ and ${\rm{BE}}_{\rmn{vol}_k}={\rm{BE}}_{\rmn{vol}}$.
			\item If $L$ is odd and $M$ is odd, then $\om = \et+\gm\wedge \xi_L$ and ${\rm{BE}}_{\rmn{vol}_k}={\rm{BE}}_{\rmn{vol}_L}$.
		\end{enumerate}
	\end{cor}
	
	Note, we extend $\gm$ by $\xi_L$ instead of just $\xi_1=\ep_{N+1}$ in case 3 only so that $\gm\wedge\xi_L$ is a $2L$-form and therefore $\om$ is homogeneous. Every choice of $k$ produces a different but equally valid Berezin integral expression. We obtain the (Pfaffian) de Bruijn integral identities for classical $\bt=1$ and $\bt=4$ when $L=1$ and $L=2$, respectively.


	\section{Statement of Results}
	\label{sec:monocharge}
	
	In all Constellation Ensembles, 
	\[
	Z_M(\vec{y})=\rmn{BE}_{\rmn{vol}}(\om(\vec{y})),
	\]
	for some appropriately defined $\om(\vec{y})$. Any time $\om(\vec{y})$ is homogeneous, we also get
	\[
	Z_M(\vec{y})=\rmn{PF}(\om(\vec{y})).
	\]
	Recall (from \autoref{ssec:setup}), in the Monocharge Constellation Ensemble, $L$ is the charge of each particle, $K$ is the number of parallel lines, and $M$ is the number of particles on each line. Let $\vec{p}$ be a complete $N$-family of monic polynomials, where $N=LKM$. Define
	\[
	\gm_L(\vec{y})=\sum_{\mft:\ul{LK}\nearrow\ul{N}}\int_{\R}\rmn{Wr}\otimes \rmn{Pr}_{\vec{y}}(\vec{p}_\mft,x)\,d\mu(x)\,\ep_{\mft},
	\]
	and define
	\[
	\et_L(\vec{y})=\sum_{\mft:\ul{LK}\nearrow \ul{N}}\sum_{\mfs:\ul{LK}\nearrow \ul{N}}\int\int_{x_1<x_2}\rmn{Wr}\otimes \rmn{Pr}_{\vec{y}}(\vec{p}_\mft,x_1)\cdot \rmn{Wr}\otimes \rmn{Pr}_{\vec{y}}(\vec{p}_\mfs,x_2)\,d\mu(x_1)\,d\mu(x_2)\,\ep_{\mft}\wedge \ep_{\mfs}.
	\]
	Provided we can write the Boltzmann factor integrand $\Om(\vec{x},\vec{y})$ as a determinant of an $N\times N$ matrix with univariate minors of the form $\rmn{Wr}\otimes \rmn{Pr}_{\vec{y}}(\vec{p}_\mft,x)$, Theorem \ref{thm:debruijngen} immediately gives us the desired Hyperpfaffian expression for the partition function $Z_M(\vec{y})$.
	
	\begin{thm}[$K$-fold Monocharge Partition Function]\label{thm:monocharge} 
		\[
		Z_M(\vec{y})=\rmn{PF}(\om(\vec{y})),
		\]
		where $\om(\vec{y})$ is defined by:
		\begin{enumerate}
			\item If $LK$ is even, then $\om(\vec{y})=\gm_L(\vec{y})$.
			\item If $LK$ is odd, but $M$ is even, then $\om(\vec{y})=\et_L(\vec{y})$. 
			\item If $LKM$ is odd, then $\om(\vec{y})=\et_L(\vec{y})+\gm_L(\vec{y})\wedge \xi_{LK}$.  
		\end{enumerate}
	\end{thm}
	As in the corollary to Theorem \ref{thm:debruijngen}, $\xi_{LK}$ upgrades $\gm_L(\vec{y})$ from an $LK$-form to a $2LK$-form and makes $\om(\vec{y})$ homogeneous so that the Hyperpfaffian $\rmn{PF}(\om(\vec{y}))$ is well-defined. Alternatively, $Z_M(\vec{y})=\rmn{BE}_{\rmn{vol}}(\om(\vec{y}))$ in cases 1 and 2, while $Z_M(\vec{y})=\rmn{BE}_{\rmn{vol}_{LK}}(\om(\vec{y}))$ in case 3.
	
	Recall also, the First Constellation Ensemble is the special case in which $L=1$. In that case, the $\rmn{Wr}\otimes \rmn{Pr}_{\vec{y}}(\vec{p}_\mft,x)$ minors are actually $\rmn{Pr}_{\vec{y}}(\vec{p}_\mft,x)$. 
	\begin{cor}[$K$-fold First Constellation Partition Function]When $L=1$, the partition function $Z_M(\vec{y})$ is given as in Theorem \ref{thm:monocharge} with the following modifications to $\gm_1(\vec{y})$ and $\et_{1}(\vec{y})$:
		\[
		\gm_1(\vec{y})=\sum_{\mft:\ul{K}\nearrow\ul{N}}\int_{\R} \rmn{Pr}_{\vec{y}}(\vec{p}_\mft,x)\,d\mu(x)\,\ep_{\mft},
		\]
		and
		\[
		\et_1(\vec{y})=\sum_{\mft:\ul{K}\nearrow \ul{N}}\sum_{\mfs:\ul{K}\nearrow \ul{N}}\int\int_{x_1<x_2} \rmn{Pr}_{\vec{y}}(\vec{p}_\mft,x_1)\cdot \rmn{Pr}_{\vec{y}}(\vec{p}_\mfs,x_2)\,d\mu(x_1)\,d\mu(x_2)\,\ep_{\mft}\wedge \ep_{\mfs}.
		\]
	\end{cor}
	
	Alternatively, any one-dimensional ensemble with a single species is a special case of a constellation ensemble in which $K=1$ (meaning only one line). Theorem \ref{thm:monocharge} agrees with Sinclair's Hyperpfaffian and Berezin integral expressions for the partition functions of $\bt$-ensembles and one-dimensional multicomponent log-gases. In particular, our $\rmn{Wr}\otimes \rmn{Pr}_{\vec{y}}(\vec{p}_\mft,x)$ minors become his $\rmn{Wr}(\vec{p}_\mft,x)$ minors when $K=1$. 
	
	To prove Theorem \ref{thm:monocharge} (and its analogues which appear in subsections \ref{ssec:hcpf}, \ref{ssec:srmulti}, and \ref{ssec:circpf}), we need to write the Boltzmann factor integrand $\Om(\vec{x},\vec{y})$ as a determinant with the appropriate structure so that Theorem \ref{thm:debruijngen} can be applied. We demonstrate this in the next section.

	\section{Confluent determinants}
	\label{sec:conf}
	Fix $\vec{L}=(L_1,\ldots L_M)\in(Z_{>0})^M$, and let $N=\sum_{m=1}^ML_m$. Let $\vec{f}=\{f_n\}_{n=1}^N$ be a family (not necessarily complete) of $\max(L_1,\ldots,L_M)-1$ times differentiable functions. Define the confluent alternant (with respect to shape $\vec{L}$) to be the $N\times N$ matrix
	\[
	V_{\vec{f}}^{\vec{L}}(\vec{x})=\begin{bmatrix}
		V_{\vec{f}}^{L_1}(x_1) & V_{\vec{f}}^{L_2}(x_2) & \cdots & V_{\vec{f}}^{L_M}(x_M)
	\end{bmatrix},
	\]
	where each $V_{\vec{f}}^{L_m}(x_m)$ is an $N\times L_m$ matrix defined by
	\[
	V_{\vec{f}}^{L_m}(x_m)=\left[D^{l-1}f_{n}(x_m)\right]_{n,l=1}^{N,L_m}.
	\]
	Then each variable $x_m$ appears in $L_m$ many consecutive columns, generated from $\vec{f}$ by taking derivatives. Note, any increasing function $\mft:\ul{L_m}\nearrow \ul{N}$ defines an $L_m\times L_m$ minor with Wronskian determinant corresponding to the polynomials $\vec{f}_\mft=\{f_{\mft(l)}\}_{l=1}^{L_m}$. Explicitly,
	\[
	\det V_{\vec{f},\mft}^{L_m}(x_m)=\rmn{Wr}(\vec{f}_\mft,x_m).
	\]
	Let $\vec{g}=\{x^{n-1}\}_{n=1}^N$. If $\vec{p}$ is any complete $N$-family of monic polynomials, then
	\[
	\det V_{\vec{g}}^{\vec{L}}(\vec{x}) = \det V_{\vec{p}}^{\vec{L}}(\vec{x})
	\]
	because $V_{\vec{p}}^{\vec{L}}(\vec{x})$ can be obtained from $V_{\vec{g}}^{\vec{L}}(\vec{x})$ by performing elementary column operations. This is only because the $p_j$ are assumed to be monic, and $\vec{p}$ is complete, containing a $p_j$ of each degree. We call $V_{\vec{g}}^{\vec{L}}(\vec{x})$ the confluent Vandermonde matrix (with respect to shape $\vec{L}$, in variables $\vec{x}$). We omit the $\vec{f}$ subscript when it is clear from context which family of functions is being used.
	
	If all $L_m$ are the same $L$, we write $V^L(\vec{x})$ for what we call the $L^{\rmn{th}}$ confluent Vandermonde matrix (in variables $\vec{x}$). Observe, the $1^{\rmn{st}}$ confluent Vandermonde matrix is the ordinary Vandermonde matrix (in $M$ many variables) whose determinant is
	\[
	\Dt(\vec{x})=\det V^1_{\vec{g}}(\vec{x})=\prod_{1\leq n<m\leq M}(x_m-x_n).
	\]
	More generally, it is known \cite{Meray1899}
	\[
	\det V_{\vec{p}}^{\vec{L}}(\vec{x})=\prod_{1\leq n<m\leq M}(x_m-x_n)^{L_mL_n}
	\]
	for any complete $N$-family of monic polynomials $\vec{p}$. In particular, 
	\[
	\det V^L_{\vec{p}}(\vec{x})=\prod_{1\leq n<m\leq M}(x_m-x_n)^{L^2}=\Dt(\vec{x})^{L^2}.
	\]
	In the previous, more general case, we will write $\Dt^{\vec{L}}(\vec{x})=\det V_{\vec{p}}^{\vec{L}}(\vec{x})$ to denote the confluent Vandermonde determinant with different exponents $L_mL_n$ on each difference $x_m-x_n$. 
	
	As an example, consider $\vec{L}=(2,3,1)$ and $\vec{g}=\{x^{n-1}\}_{n=1}^N$. For simplicity, we will use the variables $\vec{x}=(a,b,c)$. Then the three columns corresponding to $b$ are
	\[
	V^3(b)=\begin{bmatrix}
		1 & 0 & 0 \\
		b & 1 & 0 \\
		b^2 & 2b & 1 \\
		b^3 & 3b^2 & 3b \\
		b^4 & 4b^3 & 6b^2 \\
		b^5 & 5b^4 & 10b^3 \\
		b^6 & 6b^5 & 15b^4
	\end{bmatrix}.
	\]
	In the third column, we have not just the second derivative but also a denominator of $2!$. One consequence of these $l!$ denominators in $D^{l-1}$ is that we get 1's on the top diagonal. Together, the full $6\times 6$ confluent Vandermonde matrix is
	\[
	V^{\vec{L}}(\vec{x})=\begin{bmatrix}
		1 & 0 & 1 & 0 & 0 & 1 \\
		a & 1 & b & 1 & 0 & c \\
		a^2 & 2a & b^2 & 2b & 1 & c^2 \\
		a^3 & 3a^2 & b^3 & 3b^2 & 3b & c^3 \\
		a^4 & 4a^3 & b^4 & 4b^3 & 6b^2 & c^4 \\
		a^5 & 5a^4 & b^5 & 5b^4 & 10b^3 & c^5 \\
		a^6 & 6a^5 & b^6 & 6b^5 & 15b^4 & c^6 
	\end{bmatrix}.
	\]

	\subsection{Proto-confluence}
	\label{ssec:proto}
	For completeness, we will give a proof of the confluent Vandermonde determinant identity. This proof uses the following lemma:
	\begin{lem}\label{lem:ffdiff}
		Suppose $f$ is an $n$ times differentiable function, and let $\nabla_{h}^n[f](x)$ be the $n$-step finite forward difference formula for $f$ at $x$ defined by
		\[
		\nabla_h^n[f](x) = \sum_{k=0}^n(-1)^k{n\choose k}f(x+(n-k)h).
		\]
		Then
		\[
		\lim_{h\to 0}\frac{\nabla_{h}^n[f](x)}{h^n}=f^{(n)}(x).
		\]
	\end{lem}
	To prove this, it is straightforward to show by induction on $n$,
	\[
	\nabla_h^{n+1}[f](x)=\nabla_h^{n}[f](x+h)-\nabla_h^{n}[f](x),
	\]
	and then show
	\[
	\lim_{h\to 0}\frac{\nabla_h^{n}[f](x+h)-\nabla_h^{n}[f](x)}{h^{n+1}}=f^{(n+1)}(x).
	\]
	Note, this also holds for $f$ holomorphic with $x,h\in\C$. 
	
	Next, let $\vec{x}\in \R^M$, and define ${\textbf{x}}=(\textbf{x}^1,\textbf{x}^2,\ldots ,\textbf{x}^M)\in \R^{N}$ by
	\[
	\textbf{x}^{m}=(x_m,x_m+h,x_m+2h,\ldots,x_m+(L_m-1)h)\in \R^{L_m}.
	\]
	Define 
	\[
	B_{\vec{f}}^{\vec{L}}(h)=\begin{bmatrix}
		B_{\vec{f}}^{L_1}(h) & B_{\vec{f}}^{L_2}(h) & \cdots & B_{\vec{f}}^{L_M}(h)
	\end{bmatrix},
	\]
	where each $B_{\vec{f}}^{L_m}(h)$ is an $N\times L_m$ matrix defined by
	\[
	B_{\vec{f}}^{L_m}(h)=\left[\frac{\nabla_{h}^{l}[f_n](x_m)}{h^{l-1}(l-1)!}\right]_{n,l=1}^{N,L_m}.
	\]
	Note, $\nabla_{h}^l[f_n](x_m)$ is a linear combination of $f_n(x_m+(l-1)h)$ for $1\leq l\leq L_m$. Thus, by taking linear combinations of columns, 
	\[
	\det V_{\vec{f}}^1({\textbf{x}})=C_{M}^{\vec{L}}(h)\cdot \det B_{\vec{f}}^{\vec{L}}(h),
	\]
	where
	\[
	C_{M}^{\vec{L}}(h)=\prod_{m=1}^{M}\left[h^{L_m\choose 2}\prod_{l=1}^{L_m}(l-1)!\right]=\prod_{m=1}^M\Dt(h\ul{L_m}).
	\]
	By Lemma \ref{lem:ffdiff} (acting on each entry in $B_{\vec{f}}^{\vec{L}}(h)$), we have
	\[
	\det V_{\vec{f}}^{\vec{L}}(\vec{x})=\lim_{h\to 0}\det B_{\vec{f}}^{\vec{L}}(h)=\lim_{h\to 0}\frac{\det V_{\vec{f}}^1(\vec{\textbf{x}})}{C_{M}^{\vec{L}}(h)}.
	\]
	In particular, if $\vec{p}$ is a complete $N$-family of monic polynomials, then
	\begin{align*}
		\det V_{\vec{p}}^{\vec{L}}(\vec{x})&=\lim_{h\to 0}\frac{\det V_{\vec{p}}^1({\textbf{x}})}{C_{M}^{\vec{L}}(h)}\\&=\lim_{h\to 0}\frac{\left[\prod_{1\leq n<m\leq M}\left(\prod_{l=1}^{L_m-1}\prod_{k=1}^{L_n-1}(x_m-x_n+(l-k)h)\right)\right]\left[\prod_{m=1}^{M}\left(\prod_{l=1}^{L_m-1}\prod_{k=1}^{L_m-1}(l-k)h\right)\right]}{\prod_{k=1}^{M}\left[h^{L_m\choose 2}\prod_{l=1}^{L_m}(l-1)!\right]}\\&=\prod_{1\leq n<m\leq M}(x_m-x_n)^{L_mL_n}\\&=\Dt^{\vec{L}}(\vec{x}).
	\end{align*}
	Because $V_{\vec{f}}^1({\textbf{x}})$, an ordinary alternant evaluated at the translated variables ${\textbf{x}}$, gives the confluent alternant (with respect to shape $\vec{L}$) in the limit, we can call $V_{\vec{f}}^1({\textbf{x}})$ a \emph{proto-confluent} alternant (with respect to a translation vector $\vec{y}$) in the variables $\vec{x}$.

	\subsection{Proof of Theorem \ref{thm:monocharge}}
	\label{ssec:proof31}
	
	In \autoref{sec:conf}, we noted a confluent alternant has Wronskian minors. Similarly, a proto-confluent alternant has proto-Wronskian minors. Moreover, mixing these structures by feeding a translated ${\textbf{x}}$ into an already confluent alternant produces the minors at the end of \autoref{ssec:wr}. Explicitly, for $\mft:\ul{L_mK}\nearrow \ul{N}$,
	\[
	\det V^{L_m}_{\mft}(x_m)=\rmn{Wr}\otimes \rmn{Pr}_{\vec{y}}(\vec{f}_\mft,x_m)
	\]
	is an $L_mK\times L_mK$ minor of $V^{\vec{L}}({\textbf{x}})$ in the single variable $x_m$. 
	
	Define $H^{L}({\textbf{x}})$ from $V^{L}({\textbf{x}})$ by multiplying each entry by the appropriate $(-i)^{L(K-1)/2}e^{-U(x_m)}$, the $LK^{\rmn{th}}$ root of the Radon-Nikodym derivative of $\mu$. Note, there are $LK$ many columns for each variable $x_m$, so this multiplies the determinant by the $LK^{\rmn{th}}$ power of the additional factors. Using the confluent Vandermonde determinant identity,
	\[
	\det H^{L}({\textbf{x}})\,dx_1\cdots dx_m=\Dt({\textbf{x}})^{L^2}\,d\mu(x_1)\cdots d\mu(x_m)=\Om(\vec{x},\vec{y})\,dx_1\cdots dx_m.
	\]
	Thus, we have shown the joint probability density function $\Om(\vec{x},\vec{y})$ to be the determinant of a matrix with the appropriate minors which appear in \autoref{sec:monocharge}, completing the proof of Theorem \ref{thm:monocharge}.

	\section{Homogeneous constellation ensembles}
	\label{sec:homogen}
	Let $\vec{L}\in (\Z_{>0})^K$ be a vector of positive integers which we will call the \emph{charge vector} of the system. Modify the setup in \autoref{ssec:setup} by changing the charge of each $x_m+iy_k$ particle from $L$ to $L_k$. Then the $M$ many particles on each line $\R+iy_k$ all have the same charge $L_k$. The contribution of energy to the system by a charge $L_k$ particle at location $x_m+iy_k$ and a charge $L_j$ particle at location $x_n+iy_j$ is given by $-L_kL_j\log |(x_m+iy_k)-(x_n+iy_j)|$. Assuming without loss of generality $\bt=1$, the total potential energy of this new system is given by
	\begin{align*}
		E(\vec{x},\vec{y})&=\sum_{k=1}^K\sum_{m=1}^ML_kU(x_m)-\sum_{k=1}^K\sum_{n<m}^{M}L_k^2\log(x_m-x_n)-M\sum_{j<k}^{K}L_jL_k\log|i(y_k-y_j)|\\&\hspace{5mm}-\sum_{j<k}^K\sum_{n<m}^ML_jL_k\log\left((x_m-x_n)^2+(y_k-y_j)^2\right).
	\end{align*}
	Let $\textbf{L}=(\vec{L},\ldots,\vec{L})\in (\Z_{>0})^{KM}$, let $R_1=\sum_{k=1}^KL_k$, and let $R_2=\sum_{j<k}^KL_jL_k$. With this setup, the relative density of states (corresponding to varying location vectors $\vec{x}$ and translation vectors $\vec{y}$) is given by the Boltzmann factor
	\[
	\Om(\vec{x},\vec{y})=\exp(-E(\vec{x},\vec{y}))=\left|\Dt^{\textbf{L}}(\textbf{x})\right|\prod_{m=1}^{M}e^{-R_1U(x_m)}=\Dt^{\textbf{L}}(\textbf{x})\prod_{m=1}^{M}(-i)^{R_2}e^{-R_1U(x_m)}=\det H^{\textbf{L}}(\textbf{x}).
	\]
	Recall, $H^{\textbf{L}}(\textbf{x})$ was defined from $V^{\textbf{L}}(\textbf{x})$ (in \autoref{ssec:proof31}) by multiplying the entries by the Radon-Nikodym derivative of $\mu$, divided evenly over the columns. In this case, we define $d\mu(x)=(-i)^{R_2}e^{-R_1U(x)}\,dx$. Thus, with another determinantal Boltzmann factor, we can already apply Theorem \ref{thm:debruijngen} to $A(\vec{x})= H^{\textbf{L}}(\textbf{x})$.

	\subsection{Homogeneous partition functions}
	\label{ssec:hcpf}
	Recall, $H^{\textbf{L}}(\textbf{x})$ (which corresponds to shape $\textbf{L}=(\vec{L},\ldots,\vec{L})$) is the matrix which has $L_1$ many columns of derivatives evaluated at $x_1+iy_1$, $L_2$ many columns of derivatives evaluated at $x_1+iy_2$, and then so on up through $L_K$ many columns of derivatives evaluated at $x_1+iy_K$, starting over at $L_1$ many columns for $x_2+iy_1$. In general, there are $L_k$ many columns for $x_m+iy_k$, and the total $R_1=\sum_{k=1}^KL_k$ many columns corresponding to $x_m$ are consecutive. An $R_1\times R_1$ minor in $x_m$ resembles $\rmn{Wr}\otimes \rmn{Pr}_{\vec{y}}(\vec{p}_\mft,x)$ (as in the Monocharge case) but has different numbers of derivatives for each $y_k$. Define
	\[
	\rmn{Wr}^{\vec{L}} \otimes \rmn{Pr}_{\vec{y}}(\vec{f},x)=\det \left[\left[D^{l-1}f_{n}(x+iy_k)\right]_{l=1}^{L_k}\right]_{n,k=1}^{R_1,K}.
	\]
	The first column is $R_1$ many functions evaluated at $x+iy_1$. The second column is the first derivatives of those functions evaluated at the same $x+iy_1$, and so on until the first $L_1$ many columns have been exhausted. The next $L_2$ many columns are $L_2$ many derivatives of the same functions evaluated at $x+iy_2$, and so on until all $y_k$ have been exhausted. The resulting $R_1\times R_1$ matrix will have $L_k\times L_k$ Wronskian blocks evaluated at one of the $K$ many $x+iy_k$. Note,
	\[
	\lim_{\vec{y}\to 0}\frac{\rmn{Wr}^{\vec{L}}\otimes \rmn{Pr}_{\vec{y}}(\vec{f},x)}{\Dt^{\vec{L}} (i\vec{y})}=\rmn{Wr}(\vec{f},x).
	\]
	Let $\vec{p}$ be a complete $N$-family of monic polynomials, where $N=R_1M$. Define
	\[
	\gm_{\vec{L}}(\vec{y})=\sum_{\mft:\ul{R_1}\nearrow\ul{N}}\int_{\R}\rmn{Wr}^{\vec{L}}\otimes \rmn{Pr}_{\vec{y}}(\vec{p}_\mft,x)\,d\mu(x)\,\ep_{\mft},
	\]
	and define
	\[
	\et_{\vec{L}}(\vec{y})=\sum_{\mft:\ul{R_1}\nearrow \ul{N}}\sum_{\mfs:\ul{R_1}\nearrow \ul{N}}\int\int_{x_1<x_2}\rmn{Wr}^{\vec{L}}\otimes \rmn{Pr}_{\vec{y}}(\vec{p}_\mft,x_1)\cdot \rmn{Wr}^{\vec{L}}\otimes \rmn{Pr}_{\vec{y}}(\vec{p}_\mfs,x_2)\,d\mu(x_1)\,d\mu(x_2)\,\ep_{\mft}\wedge \ep_{\mfs}.
	\]
	Applying Theorem \ref{thm:debruijngen} in this context produces the following generalization of Theorem \ref{thm:monocharge}: 
	\begin{thm}[$K$-fold Homogeneous Partition Function]\label{thm:homogen}
		\[
		Z_M(\vec{y})=\int_{-\I<x_1<\ldots<x_M<\I}\Om(\vec{x},\vec{y})\,dx_1\cdots dx_M=\rmn{PF}(\om(\vec{y})),
		\]
		where $\om(\vec{y})$ is defined by:
		\begin{enumerate}
			\item If $R_1$ is even, then $\om(\vec{y})=\gm_{\vec{L}}(\vec{y})$.
			\item If $R_1$ is odd, but $M$ is even, then $\om(\vec{y})=\et_{\vec{L}}(\vec{y})$. 
			\item If $R_1M$ is odd, then $\om(\vec{y})=\et_{\vec{L}}(\vec{y})+\gm_{\vec{L}}(\vec{y})\wedge \xi_{R_1}$. 
		\end{enumerate}
	\end{thm}
	The three cases are the same as those appearing in Theorem \ref{thm:monocharge}, replacing all instances of $LK$ with $R_1=\sum_{k=1}^{K}L_k$. As before, the $\xi_{R_1}$ in case 3 is a pure tensor which upgrades $\gm_{\vec{L}}(\vec{y})$ from an odd $R_1$-form to an even $2R_1$-form so that the Hyperpfaffian is well-defined. As mentioned in \autoref{sec:intro}, Homogeneous Constellation Ensembles are the most general classification (in this volume) for which the partition functions are Hyperpfaffians (because of homogeneous $\om(\vec{y})$).

	\section{Limits of linear constellations}
	\label{sec:limlin}

	Starting with a Homogeneous Constellation Ensemble, taking the limit as $\vec{y}\to 0$ produces infinite potential energy, so the resulting Boltzmann factor $\Om_M(\vec{x},0)$ is identically zero. In our physical interpretation, collapsing the parallel lines onto each other forces particles with the same real parts (who want to repel each other) onto each other. This is represented by the interaction terms with $L_jL_k\log|i(y_k-y_j)|$. To obtain meaningful limits, we remove these singularities by removing the appropriate interaction terms. Taking the limit inside the integral, it is easy to see
	\[
	\lim_{\vec{y}\to 0}\frac{\Dt^{\textbf{L}}(\textbf{x})}{\left(\Dt^{\vec{L}}(i\vec{y})\right)^M}=\Dt(\vec{x})^{R_1^2}.
	\]
	Thus, the limiting Boltzmann factor corresponds to a one-dimensional ensemble of particles with charge $R_1=\sum_{k=1}^KL_k$. In terms of confluent matrices,
	\[
	\lim_{\vec{y}\to 0}\frac{V^{\textbf{L}}(\textbf{x})}{\left(\Dt^{\vec{L}}(\vec{iy})\right)^M}=V^{R_1}(\vec{x}).
	\]
	This limit turns all proto-confluent translation columns into further derivative columns, a total of $R_1$ for each variable $x_m$. In terms of the partition function,
	\begin{align*}
		\lim_{\vec{y}\to 0}\frac{Z_{M}(\vec{y})}{\left(\Dt^{\vec{L}}(i\vec{y})\right)^M}&=\lim_{\vec{y}\to 0}\int\frac{\gm_{\vec{L}}(\vec{y})^{\wedge M}}{M!\left(\Dt^{\vec{L}}(i\vec{y})\right)^M}\ep_{\rmn{vol}}\\&=\lim_{\vec{y}\to 0}\frac{1}{M!}\int\left[\sum_{\mft:\ul{R_1}\nearrow\ul{N}}\frac{1}{\Dt^{\vec{L}}(i\vec{y})}\int_{\R}\rmn{Wr}^{\vec{L}}\otimes \rmn{Pr}_{\vec{y}}(\vec{p}_\mft,x)\,d\mu(x)\,\ep_{\mft}\right]^{\wedge M}\ep_{\rmn{vol}}\\&=\frac{1}{M!}\int \left[\sum_{\mft:\ul{R_1}\nearrow\ul{N}}\int_{\R}\rmn{Wr}(\vec{p}_\mft,x)\,d\mu(x)\,\ep_{\mft}\right]^{\wedge M}\ep_{\rmn{vol}},
	\end{align*}
	in the case $R_1K$ is even. The limit of the Hyperpfaffian is again Hyperpfaffian. Also, the Wronskian minors which appear in this Hyperpfaffian are the minors of the confluent limit of the proto-confluent matrix $V^{\textbf{L}}(\textbf{x})$. An analogous result holds for $R_1K$ odd, attaching a $\Dt^{\vec{L}}(i\vec{y})$ denominator to each of two Wronskian-like integrands at a time. 
	
	It is not necessary that all $y_k$ go to zero. We could instead take limits as some $y_j\to y_k$. Physically, this means collapsing some lines together but not all. If we didn't already have the confluent Vandermonde technology, we could produce any Homogeneous Constellation Ensemble as a limit of First Linear Constellation Ensembles (in which case all the particles have charge 1 and only the ordinary Vandermonde determinant is needed). As a special case of this, collapsing $K$ many lines produces a one-dimensional $\beta=K^2$ ensemble (of charge $K$ particles).

	\subsection{Limits at infinity}
	\label{ssec:liminf}
	Next, we consider limits as the distances between our lines increase without bound. Not only do we want $y_k\to \I$, but also $(y_k-y_j)\to \I$. For simplicity, we start by setting $y_k=(k-1)h$ and then consider limits as $h\to \I$. This limit produces negatively infinite potential energy, so the resulting Boltzmann factor is positively infinite. This comes from interaction terms with $L_jL_k\log\left((x_m-x_n)^2+(y_k-y_j)^2\right)$. Denote
	\[
	G^{\vec{L}}_{M}(h)=\left[\prod_{j<k}\left(1+((k-j)h)^2\right)^{L_jL_k}\right]^{M\choose 2}.
	\]
	Note, $\ds\lim_{h\to 0}G^{\vec{L}}_{M}(h)=1$, so we can add $G^{\vec{L}}_{M}(h)$ to the denominators in \autoref{sec:limlin} without changing the limits (as $h\to 0$). On the other hand, it is straightforward to check
	\[
	\lim_{h\to \I}\frac{\Dt^{\textbf{L}}(\textbf{x})}{\left(\Dt^{\vec{L}}(ih\ul{K})\right)^MG^{\vec{L}}_{M}(h)}=\Dt(\vec{x})^{L_1^2+\cdots+L_K^2}.
	\]
	\hspace{5mm}Thus, in terms of the Boltzmann factor, the limit produces a one-dimensional $\bt=\sum_{k}L_k^2$ ensemble (of charge $\sqrt{\sum_{k}L_k^2}$). As a special case of this, if we take the limit of the First Linear Constellation, the result is a one-dimensional $\bt=K$ ensemble corresponding to possibly non-integer charge $\sqrt{K}$. Physically, moving our lines away from each other without bound breaks the interactions between particles from different lines. The remaining energy contributions from internal interactions within each line are additive. A pair of charge one particles repel another pair of charge one particles with a force greater than that between just two charge 1 particles but weaker than that of two charge 2 particles. Together with \autoref{sec:limlin}, we now have an interpolation between one-dimensional $\bt=K^2$ and $\bt=K$ ensembles. 
	
	In terms of confluent matrices, our existing methods do not allow us to produce square-free powers of the ordinary Vandermonde determinant. Additionally, it is unclear how to bring the limit inside $V^{\textbf{L}}(\textbf{x})$ in hopes of producing an entirely new determinantal expression for $\Dt(\vec{x})^{L_1^2+\cdots+L_K^2}$, which as stated is a power of a determinant, not a lone determinant. Equivalently, it is unclear how to distribute the denominator of the limit over the Wronskian-like minors of the confluent determinant (which would have allowed us to bring the limit inside the Hyperpfaffian expression for the partition functions). Without this, the limit of the Hyperpfaffian partition function cannot simply be written as a Hyperpfaffian using the methods demonstrated thus far (from Theorem \ref{thm:debruijngen}). However, for each $h$ (or $\vec{y}$) fixed along the way, the partition function is Hyperpfaffian as stated in Theorem \ref{thm:homogen}. 
	
	Recall (from \autoref{sec:hpf}), Shum considered the $2$-fold First Constellation Ensembles in his 2013 dissertation. First, he demonstrated the partition function is Pfaffian (instead of Hyperpfaffian, because $K=2$). Using this, he gave the kernel of which the correlation functions are the Pfaffian. When computing the limits (as $h\to 0$ and $h\to \I$), he worked directly with the kernel, producing the expected kernels of the limiting ensembles in both directions (classical $\bt=4$ as $h\to 0$ and classical $\bt=2$ as $h\to \I$). In this way, the limiting ensembles were demonstrated to be solvable Pfaffian point processes without needing to explicitly express the limiting partition functions as Pfaffians. Analogously, square-free $\bt=K$ ensembles may still have Hyperpfaffian correlation functions even though the methods of this volume do not produce an explicitly Hyperpfaffian partition function in the limit.

	\section{Multicomponent constellation ensembles}
	\label{sec:multi}
	In \autoref{ssec:setup}, we demonstrated the Boltzmann factor of the Monocharge Constellation is the same as the Boltzmann factor of the single-species $\bt$-ensemble with $\bt=L^2$ but with the $KM$ many translated variables $\textbf{x}$ substituted in. In both cases, the Boltzmann factors are determinantal. The Wronskian minors of the former resemble the minors of the latter except made proto-confluent by the addition of the translation vector $\vec{y}$. Likewise, the forms which give the partition functions for Multicomponent Constellation Ensembles are simply the proto-confluent versions of the forms which give the partition functions for one-dimensional multicomponent log-gases. 
	
	In a multicomponent log-gas, we say particles with the same charge belong to the same \emph{species}, and we assume they are indistinguishable. In the previous volume, we considered two ensembles:
	\begin{enumerate}
		\item \emph{The Canonical Ensemble}, in which the number of particles of each species is fixed; in this case, we say fixed population.
		\item \emph{The Isocharge Grand Canonical Ensemble}, in which the sum of the charges of the particles is fixed, but the number of particles of each species is allowed to vary; in this case we say the total charge of the system is fixed.
	\end{enumerate}
	
	In contrast, the \emph{Grand Canonical Ensemble} traditionally refers to the ensemble in which the total number of particles is not fixed. For computational purposes, it is beneficial to group configurations which share the same total charge. The Grand Canonical Ensemble is then a disjoint union (over all possible sums of charges) of our Isocharge ensembles. By first conditioning on the number of particles of each species, the partition function for the Isocharge Grand Canonical Ensemble is built up from the partition functions of the Canonical type, revealing the former to be a generating function of the latter as a function of the \emph{fugacities} of each species (roughly, the probability of the occurrence of any one particle of a given charge). Likewise, we will consider analogous versions for our Constellation Ensembles. For reference, we will start by giving abbreviated versions of the setup and the main results from the one-dimensional case.

	\subsection{One-dimensional setup}
	\label{ssec:1dim}
	Let $J\in \Z_{>0}$ be a positive integer, the maximum number of distinct charges in the system. Let $\vec{L}=(L_1,\ldots,L_J)\in (\Z_{>0})^J$ be a vector of distinct positive integers which we will call the \emph{charge vector} of the system. We will assume the first $r$ many $L_j$ are even, and the remaining $J-r$ many are odd. Let $\vec{M}\in (\Z_{\geq 0})^J$ be a vector of non-negative integers which we will call the \emph{population vector} of the system. Each $M_j$ gives the number (possibly zero) of indistinguishable particles of charge $L_j$ on the real axis. Let $N=\vec{L}\cdot \vec{M}$ be the total charge of the system, and let $U:\R\to \R$ be a potential on the real axis. 
	
	For any complete $N$-family of monic polynomials $\vec{p}$, define
	\[
	\gm_j=\sum_{\mft:\ul{L_j}\nearrow\ul{N}}\int_{\R}\rmn{Wr}(\vec{p}_\mft,x)\,d\mu_j(x)\,\ep_{\mft},
	\]
	and define
	\[
	\et_{j,k}=\sum_{\mft:\underline{L_j}\nearrow\underline{N}}\sum_{\mathfrak{s}:\underline{L_k}\nearrow\underline{N}}\int\int_{x<y}\rmn{Wr}(\vec{p}_{\mft},x)\rmn{Wr}(\vec{p}_{\mathfrak{s}},y)\,d\mu_j(x)d\mu_k(y)\,\ep_{\mft}\wedge\ep_{\mathfrak{s}},
	\]
	where $d\mu_j(x)=\exp(-L_jU(x))\,dx$.

	\subsection{Canonical ensemble}
	\label{ssec:canon}
	Under the setup in \autoref{ssec:1dim}, we inherit the following lemma from our previous volume:
	\begin{lem}[Canonical Partition Function]\label{lem:canpf}
		When $N$ is even,
		\[
		Z_{\vec{M}}=\int\frac{\gm_{1}^{\wedge M_1}}{M_1!}\wedge \cdots \wedge \frac{\gm_{r}^{\wedge M_r}}{M_r!}\wedge \sum_{\sm\in \rmn{Sh}(M_{r+1},\ldots,M_J)}\frac{1}{K!}\bigwedge_{j=1}^J\bigwedge_{k=1}^J\et_{j,k}^{\wedge M_{\sm}^{j,k}}\,d\ep_{\rmn{vol}},
		\]
		and when $N$ is odd,
		\[
		Z_{\vec{M}}=\int\frac{\gm_{1}^{\wedge M_1}}{M_1!}\wedge \cdots \wedge \frac{\gm_{r}^{\wedge M_r}}{M_r!}\wedge \sum_{\sm\in \rmn{Sh}(M_{r+1},\ldots,M_J)}\frac{1}{K!}\bigwedge_{j=1}^J\bigwedge_{k=1}^J\et_{j,k}^{\wedge M_{\sm}^{j,k}}\wedge \gm_{\sm^{-1}(K_o)}\,d\ep_{\rmn{vol}}.
		\]
		Here, $K_o=\sum_{j=r+1}^{J}M_j$.  $\rmn{Sh}(M_{r+1},\ldots,M_J)\subset S_{K_o}$ denotes the subset of shuffle permutations. These are permutations which satisfy $\sm(n)<\sm(m)$ whenever
		\[
		M_{r+1}+M_{r+2}+\cdots +M_{j}<n<m\leq M_{r+1}+M_{r+2}+\cdots +M_{j+1},
		\]
		meaning each block of $M_j$ many things retains the same relative order in the image of $\sm$. If we define 
		\[
		\vec{\ld}=(L_{r+1},\ldots,L_{r+1},L_{r+2},\ldots, L_J,\ldots,L_{J}),
		\]
		which has $M_j$ copies of each $L_j$ for $r+1\leq j\leq J$, then each $M_\sm^{j,k}$ is the number of times $\ld_{\sm^{-1}(2m-1)}=L_j$ and $\ld_{\sm^{-1}(2m)}=L_k$. When $N$ is even, $K=\sum_{j,k}M_\sm^{j,k}$. When $N$ is odd, $K=1+\sum_{j,k}M_\sm^{j,k}$. 
	\end{lem}
	Heuristically, the sum is taken over all possible orderings of the particles which carry an odd charge $L_j$. Since particles of the same charge are indistinguishable, we restrict the sum to shuffle permutations which preserve the relative order of blocks of $M_j$ many particles. $K_o$ is the total number of these particles with odd charges. $\vec{\ld}$ is the vector of charges, each of which is repeated $M_j$ many times. Pairs $(L_j,L_k)$ of adjacent charges produce the double-Wronskian $\et_{j,k}$ form by combining an $L_j\times L_j$ Wronskian in one variable with an $L_k\times L_k$ Wronskian in another variable. Because the charges are repeated, a permutation $\sm$ may produce multiple copies of the same pair $(L_j,L_k)$ and multiple copies of the same $\et_{j,k}$. $K$ is the number of these pairs, also the number of $\et_{j,k}$ factors in the wedge product.  When $N$ is odd, the last particle lacks a partner and produces only a single-Wronskian $\gm_j$ form. 
	
	As an example, consider four particles of even charge $2$ at locations $a_1,a_2,a_3,a_4$, two particles of odd charge $3$ at locations $b_1,b_2$, and three particles of odd charge $5$ at locations $c_1,c_2,c_3$. Then $\vec{L}=(2,3,5)$, $\vec{M}=(4,2,3)$, and $N=8+6+15=29$. Regardless of permutation on the particles of odd charge, the even charge 2 particles produce 4 copies of the single-Wronskian 2-form $\gm_1$, built from $2\times 2$ Wronskians. 
	
	Under the identity permutation which orders the odd variables as $b_1,b_2,c_1,c_2,c_3$, we pair $3\times 3$ Wronskians in $b_1$ with  $3\times 3$ Wronskians in $b_2$ to produce the 6-form $\et_{2,2}.$ Similarly, pairing $5\times 5$ Wronskians in variable $c_1$ with $5\times 5$ Wronskians in variable $c_2$ produces the 10-form $\et_{3,3}$. The last variable $c_3$ is unpaired because we have an odd number of variables. This unpaired $c_3$ produces the 5-form $\gm_3$. Thus, the wedge product corresponding to the identity permutation is 
	\[
	\frac{\gm_1^{\wedge 4}}{4!}\wedge \frac{\et_{2,2}\wedge \et_{3,3}\wedge \gm_3}{3!}\in \bigwedge(\R^{29}).
	\]
	Next, let $\sm$ be the permutation which swaps $b_2$ with $c_1$. Under this permutation, we pair $b_1$ with $c_1$ and then $b_2$ with $c_2$. The result is two copies of the 8-form $\et_{2,3}$. Again, $c_3$ is still left unpaired. The corresponding wedge product is
	\[
	\frac{\gm_1^{\wedge 4}}{4!}\wedge \frac{\et_{2,3}^{\wedge2}\wedge \gm_3}{3!}\in \bigwedge(\R^{29}).
	\]
	The permutation which swaps $b_1$ with $b_2$ produces the same $\et_{2,2}$ (and $\et_{3,3}$) as the identity permutation. To avoid this redundancy, we consider only shuffle permutations. The permutation which moves $c_1$ to the front (ordering the variables as $c_1,b_1,b_2,c_2,c_3$) produces $\et_{3,2}$ followed by the distinct $\et_{2,3}$. 
	
	Next, consider the permutation which puts all three of the charge 5 particles before the two charge 3 particles. We pair a charge 5 with a charge 5, pair the last charge 5 with a charge 3, and leave a charge 3 unpaired. The corresponding wedge product is
	\[
	\frac{\gm_1^{\wedge 4}}{4!}\wedge \frac{\et_{3,3}\wedge \et_{3,2}\wedge \gm_2}{3!}\in \bigwedge(\R^{29}).
	\]

	\subsection{Isocharge grand canonical ensemble}
	\label{ssec:isogrand}
	Allowing the number of particles of each species to vary, let $P(\vec{M})$ be the probability of finding the system with population vector $\vec{M}$. Let $\vec{z}=(z_1,\ldots,z_J)\in (\R_{>0})^J$ be a vector of positive real numbers called the \emph{fugacity vector}. Classically, the probability $P(\vec{M})$ is given by
	\[
	P(\vec{M})=z_1^{M_1}z_2^{M_2}\cdots z_{J}^{M_J}\frac{Z_{\vec{M}}}{Z_N},
	\]
	where $Z_N$ is the partition function of the Isocharge Grand Canonical Ensemble (corresponding to fixed total charge $N$) given by
	\[
	Z_N=\sum_{\vec{L}\cdot \vec{M}=N}z_1^{M_1}z_2^{M_2}\cdots z_{J}^{M_J}{Z_{\vec{M}}}.
	\]
	In the above expression, the vector $\vec{L}$ of allowed charges is fixed, so we're summing over allowed population vectors $\vec{M}$. A population vector is valid only when the sum of the charges $\sum_{j=1}^JL_jM_j$ is equal to the prescribed total charge $N$. Taking the fugacity vector $\vec{z}$ to be a vector of indeterminants, $Z_N$ is a polynomial in these indeterminants which generates the partition functions of the Canonical Ensembles. Combining partition functions of the Canonical type, we obtained the following theorem:
	\begin{thm}[Isocharge Grand Canonical Partition Function]\label{thm:isogcpf}
		When $N$ is even,
		\[
		Z_N=\rmn{BE}_{\rmn{vol}}\left(\sum_{j=1}^rz_j\gm_j+\sum_{j=r+1}^{J}\sum_{k=r+1}^{J}z_jz_k\et_{j,k}\right),
		\]
		and when $N$ is odd,
		\[
		Z_N=\rmn{BE}_{\rmn{vol}}\left(\sum_{j=1}^rz_j\gm_j+\sum_{j=r+1}^{J}\sum_{k=r+1}^{J}z_jz_k\et_{j,k}+\sum_{j=r+1}^{J}z_j\gm_j\wedge \ep_{N+1}\right).
		\]
	\end{thm}
	Note, this is the first case (in this volume) in which the partition function is not Hyperpfaffian (because the form inside the parentheses is not homogeneous).

	\subsection{Constellation partition functions}
	\label{ssec:constpf}
	Starting with a one-dimensional configuration on the line $\R+iy_1$, copy the configuration onto the other lines $\R+iy_k$ for $K$ many total copies of the same one-dimensional configuration (following the procedure outlined in \autoref{ssec:setup}). With this setup, we can take Lemma \ref{lem:canpf} and Theorem \ref{thm:isogcpf} entirely as written with only slight modification to how $\gm_j$ and $\et_{j,k}$ are defined (to account for the added translation vector $\vec{y}$). 
	
	Let $\textbf{x}=(\textbf{x}^1,\ldots, \textbf{x}^J)\in \R^{M_1}\times \cdots \times \R^{M_J}$ so that $\textbf{x}^{j}=(x_1^j,\ldots,x_{M_j}^j)\in \R^{M_j}$ gives the real parts of all particles of charge $L_j$. Define $\textbf{x}_{\vec{y}}=(\textbf{x}^1_{\vec{y}},\ldots, \textbf{x}^J_{\vec{y}})\in \C^{KM_1}\times \cdots \times \C^{KM_J}$ so that
	\[
	\textbf{x}^j_{\vec{y}}=(x_1^j+i{\vec{y}},\ldots,x_{M_j}^j+i{\vec{y}})\in \C^{KM_j},
	\]
	with $x_m^j+i\vec{y}=(x_m^j+iy_1,\ldots,x_m^j+iy_k)\in \C^K$. As a list, $\textbf{x}_{\vec{y}}$ is generated from $\textbf{x}$ by replacing each real location $x_m^j$ with $x_m^j+i\vec{y}$, the list of its $K$ many translations. Let $N=K(\vec{L}\cdot \vec{M})$ be the total charge of this expanded system. 
	
	Without writing out the full Boltzmann factor for the interactions between these particles, it is straightforward to verify all instances of $i$ vanish (as in \autoref{ssec:setup} and the analogous start of \autoref{sec:homogen}) except for the interactions between two particles which share a real part. To obtain the absolute value of these factors, we factored out powers of $-i$ and included them in the (complex) measure $\mu$ (which otherwise comes just from the potential $U$). Dealing with one real part at a time, we can use what we know from the Monocharge case to get the correct combinatorial exponent on $-i$.
	
	Explicitly, for any real part $x_m^j$ (corresponding to a constellation of $K$ many charge $L_j$ particles), the energy contribution from the potential is $L_jU(x_m^j)$ times the number of translations $K$. We get one factor of $i^{L_j^2}$ for each pair of particles in the constellation $x_m^j+i\vec{y}$ of which there are ${K\choose 2}$ many. Thus, we set 
	\[
	d\mu_j(x)=\left((-i)^{L_j(K-1)/2}e^{-U(x)}\right)^{L_jK}dx.
	\]
	With any Homogeneous Constellation Ensemble, we could assume the real parts were ordered $x_1<\ldots<x_M$ because all particles on the same line were indistinguishable (same charge). This was necessary to drop the absolute value from the Boltzmann factor. In particular, whenever the real parts are labeled with the same order as the domain of integration, all differences in the confluent Vandermonde determinant are positive. In the case of differently-charged particles, the order in which they occur is relevant, and some additional tools are needed.
	
	For $\sm\in S_n$, let $Q_n(\sm)\subset \R^n$ denote the region with $x_{\sm^{-1}(1)}<\cdots<x_{\sm^{-1}(n)}$. The sign of the confluent Vandermonde determinant is constant on this region of $\R^n$. Taking the real absolute value of the confluent Vandermonde determinant equates to replacing the matrix with one in which the columns have been reordered to match $\sm$ (so that all differences in the determinant are positive on $Q_n(\sm)$). 
	
	Given confluent type matrix $V^{\vec{L}}(\vec{x})$, let $V^{\vec{L}}_{\sm}(\vec{x})$ denote the matrix obtained from $V^{\vec{L}}(\vec{x})$ by permuting the columns so that the columns with $x_{\sm^{-1}(1)}$ come first (of which there are $L_{\sm^{-1}(1)}$ many), then the columns with $x_{\sm^{-1}(2)}$ come next (of which there are $L_{\sm^{-1}(2)}$ many), and so on until the $x_j$ are exhausted. Then for $\vec{x}\in Q_n(\sm)$, we have $|\det V^{\vec{L}}(\vec{x})|=\det V^{\vec{L}}_{\sm}(\vec{x})$. 
	
	Let $\Ld=(L_1,\ldots,L_1,L_2,\ldots,L_2,\ldots,L_J,\ldots,L_J)$ with each $L_j$ appearing $KM_j$ consecutive times. Then, the partition function for the Canonical Multicomponent Constellation Ensemble (meaning fixed population $\vec{M}$) is given by
	\begin{align*}
		Z_{\vec{M}}(\vec{y})&=\frac{1}{M_1!\cdots M_J!}\sum_{\sm \in S_{T}}\int_{Q_T(\sm)}\det H^{\Ld}_\sm(\textbf{x}_{\vec{y}})\,d(\textbf{x}),
	\end{align*}
	where $T=\sum_{j=1}^{J}KM_j$. Note, the factorial denominators appear because there are $M_j$ indistinguishable real parts which share the same charge $L_j$. This same technique (of decomposing the domain of integration over the possible orderings on the variables to account for the absolute value) is also necessary in the one-dimensional case, so nothing unique to Constellation Ensembles is happening here. This is where the sum in Lemma \ref{lem:canpf} comes from.

	\subsection{Statement of results}
	\label{ssec:srmulti}
	Writing the integrand as a determinant without absolute value was the last step needed to write the partition function as the Berezin integral of an exponential (as prescribed by Lemma \ref{lem:canpf} and Theorem \ref{thm:isogcpf}). We define $\gm_j(\vec{y})$ and $\et_{j,k}(\vec{y})$ by taking appropriate univariate minors of $H^\Ld(\textbf{x}_{\vec{y}})$ of different sizes. Define
	\[
	\gm_j(\vec{y})=\sum_{\mft:\ul{L_jK}\nearrow \ul{N}}\int_{\R}\rmn{Wr}\otimes \rmn{Pr}_{\vec{y}}(\vec{p}_\mft,x)\,d\mu(x)\,\ep_{\mft},
	\]
	and define
	\[
	\et_{j,k}(\vec{y})=\sum_{\mft:\ul{L_jK}\nearrow \ul{N}}\sum_{\mfs:\ul{L_kK}\nearrow \ul{N}}\int\int_{x_1<x_2}\rmn{Wr}\otimes \rmn{Pr}_{\vec{y}}(\vec{p}_\mft,x_1)\cdot \rmn{Wr}\otimes \rmn{Pr}_{\vec{y}}(\vec{p}_\mfs,x_2)\,d\mu(x_1)\,d\mu(x_2)\,\ep_{\mft}\wedge \ep_{\mfs}.
	\]
	When $K$ is odd, we're done (because the parity of $L_jK$ is determined by the $L_j$). However, when $K$ is even, all of our minors have even dimensions $L_jK$, and the total charge $N=K(\vec{L}\cdot \vec{M})$ is even as well. Thus, for $K$ even (and no additional restrictions on $L_j$), we get the ``all even" versions of our Berezin integral expressions:
	\begin{thm}\label{thm:allevens}
		When $K$ is even,
		\[
		Z_{\vec{M}}(\vec{y})=\int\frac{\gm_1(\vec{y})^{\wedge M_1}}{M_1!}\wedge\cdots\wedge\frac{\gm_J(\vec{y})^{\wedge M_J}}{M_J!}\,d\ep_{\rmn{vol}},
		\]
		and
		\[
		Z_{N}(\vec{y})=\rmn{BE}_{\rmn{vol}}\left(\sum_{j=1}^Jz_j\gm_j(\vec{y})\right).
		\]
	\end{thm}
	Note, the fugacity parameters $z_j$ are no longer the probabilities of individual charge $L_j$ particles appearing. Instead, $z_j$ is the probability of a constellation of $K$ many points all having the same charge $L_j$. 
	
	\section{Circular ensembles}
	\label{sec:circ}
	We will begin with Homogeneous Circular Constellation Ensembles of which Monocharge Circular Constellation Ensembles are a special case. Consider $K$ concentric circles in the complex plane with radii $\vec{y}$. Define $\vec{L}$ and $\textbf{L}=(\vec{L},\ldots,\vec{L})\in (\N)^{KM}$ as in \autoref{sec:homogen}. Replace $\R$ in the definition of $\vec{x}$ by $[0,2\pi)$. For each angle $x_m\in [0,2\pi)$, and $1\leq k\leq K$, place a charge $L_k$ particle at location $y_ke^{ix_m}$. Denote the (total $KM$) particle locations by 
	\[
	\textbf{z}=(\textbf{z}^1,\textbf{z}^2,\ldots,\textbf{z}^M)\in \C^{KM}
	\]
	where $\textbf{z}^m=\vec{y}e^{ix_m}=(y_1e^{ix_m},y_2e^{ix_m},\ldots,y_Ke^{ix_m})\in \C^K$. Assuming logarithmic interaction between the particles, the total potential energy of this system is given by
	\begin{align*}
		E(\vec{x},\vec{y})&=-\sum_{k=1}^K\sum_{n<m}^ML_k^2\log\left|y_ke^{ix_m}-y_ke^{ix_n}\right|-\sum_{j<k}^{K}\sum_{m=1}^ML_jL_k\log\left|y_ke^{ix_m}-y_je^{ix_m}\right|\\&\hspace{5mm}-\sum_{n<m}^M\sum_{j<k}^KL_jL_k\log\left|y_ke^{ix_m}-y_je^{ix_n}\right|+L_jL_k\log\left|y_je^{ix_m}-y_ke^{ix_n}\right|.
	\end{align*}
	As observed in \cite{Mehta2004}, we can express the Boltzmann factor $\Om_M(\vec{x},\vec{y})=e^{-E(\vec{x},\vec{y})}$ without absolute values using the following identities:
	\[
	\left|y_ke^{ix_m}-y_ke^{ix_n}\right|=-ie^{-i(x_m+x_n)/2}\left(y_ke^{ix_m}-y_ke^{ix_n}\right)\sgn(x_m-x_n).
	\]
	\[
	\left|y_ke^{ix_m}-y_je^{ix_m}\right|=e^{-ix_m}\left(y_ke^{ix_m}-y_je^{ix_m}\right).
	\]
	\[
	\left|y_ke^{ix_m}-y_je^{ix_n}\right|\left|y_je^{ix_m}-y_ke^{ix_n}\right|=-e^{-i(x_m+x_n)}\left(y_ke^{ix_m}-y_je^{ix_n}\right)\left(y_je^{ix_m}-y_ke^{ix_n}\right).
	\]
	As in \autoref{ssec:setup} and \autoref{sec:homogen}, we can assume without loss of generality $\vec{x}\in Q_M(\id)$, meaning $x_1<\cdots< x_M$. Then $\sgn(x_m-x_n)>0$ for all $n<m$. Thus, the relative density of states (corresponding to varying location vectors $\vec{x}$ and translation vectors $\vec{y}$) is given by the Boltzmann factor
	\[
	\Om(\vec{x},\vec{y})=e^{-E(\vec{x},\vec{y})}=\left|\Dt^{\textbf{L}}(\textbf{z})\right|=\Dt^{\textbf{L}}(\textbf{z})\prod_{m=1}^M(-ie^{-ix_m})^{R_3(M-1)/2}(e^{-ix_m})^{R_2}=\det H^{\textbf{L}}(\textbf{z}),
	\]
	where $R_2=\sum_{j<k}^KL_jL_k$, $R_3=\sum_{j,k=1}^KL_jL_k$, and $d\mu(x)=(-ie^{-ix})^{R_3(M-1)/2}(e^{-ix})^{R_2}\,dx.$

	\subsection{Circular partition functions}
	\label{ssec:circpf}
	Recall from \autoref{ssec:hcpf}, $R_1=\sum_{k=1}^KL_k$. For Circular Constellation Ensembles, instead of $L_k$ columns for each of the $x_m+iy_k$ (in the linear case), $H^{\textbf{L}}(\textbf{z})$ has $L_k$ columns for each of the $y_ke^{ix_m}$. Define
	\[
	\rmn{Cr}_{\vec{y}}(\vec{f},x)=\det[f_{n}(y_ke^{ix})]_{n,k=1}^K,
	\]
	and define 
	\[
	\rmn{Wr}^{\vec{L}}\otimes \rmn{Cr}_{\vec{y}}(\vec{f},x)=\det \left[\left[D^{l-1}f_{n}(y_ke^{ix})\right]_{l=1}^{L_k}\right]_{n,k=1}^{R_1,K}.
	\]
	This is analogous to the definition of $\rmn{Wr}^{\vec{L}}\otimes \rmn{Pr}_{\vec{y}}(\vec{f},x)$ with linear translations $x+iy_k$ replaced with circular translations $y_ke^{ix}$. These are the $R_1\times R_1$ minors of $H^{\textbf{L}}(\textbf{z})$ which correspond to a single position $x_m$. 
	
	Proceeding as in \autoref{ssec:hcpf}, define
	\[
	\gm_{\vec{L}}(\vec{y})=\sum_{\mft:\ul{R_1}\nearrow\ul{N}}\left[\int_{0}^{2\pi}\rmn{Wr}^{\vec{L}}\otimes \rmn{Cr}_{\vec{y}}(\vec{p}_\mft,x)\,d\mu(x)\right]\ep_{\mft},
	\]
	and define
	\[
	\et_{\vec{L}}(\vec{y})=\sum_{\mft:\ul{R_1}\nearrow \ul{N}}\sum_{\mfs:\ul{R_1}\nearrow \ul{N}}\left[\int\int_{0<x_1<x_2<2\pi}\rmn{Wr}^{\vec{L}}\otimes \rmn{Cr}_{\vec{y}}(\vec{p}_\mft,x_1)\cdot \rmn{Wr}^{\vec{L}}\otimes \rmn{Cr}_{\vec{y}}(\vec{p}_\mfs,x_2)\,d\mu(x_1)\,d\mu(x_2)\right]\ep_{\mft}\wedge \ep_{\mfs}.
	\]
	By Theorem \ref{thm:debruijngen}, the expressions for partition functions of Homogeneous Circular Ensembles are the same as in Theorem \ref{thm:homogen} using these new (circular) $\gm_{\vec{L}}(\vec{y})$ and $\et_{\vec{L}}(\vec{y})$. For Monocharge Constellation Ensembles, we can specialize to the expressions given in Theorem \ref{thm:monocharge}.

	\subsection{Computational techniques}
	\label{ssec:comp}
	Recall (from \autoref{sec:conf}), $\det V^{\vec{L}}_{\vec{p}}(\vec{x})=\Dt^{\vec{L}}(\vec{x})$ for any choice of complete $N$-family of monic polynomials $\vec{p}$. When actually computing integrals of Wronskians, some choices are better than others. In some cases (such as the circular case), there exists polynomials for which the integrals of Wronskians are often zero. This depends on the measures $\mu$ which come from the potential $U$. 
	
	For example, consider $\vec{g}=\{x^{n-1}\}_{n=1}^N$. It is straightforward to verify that the Wronskian of a collection of monomials will again be a monomial. In particular, for any $\mft:\ul{K}\nearrow\ul{N}$, we have
	\[
	\rmn{Wr}(\vec{g}_{\mft},x)=x^{\sum_k\mft(k)-k}\frac{\Dt(\mft(\ul{K}))}{\Dt(\ul{K})}.
	\]
	Thus, for $R\in \Z$,
	\[
	\int_{0}^{2\pi}\rmn{Wr}(\vec{g}_{\mft},re^{ix})(e^{-ix})^{R}\,dx=\int_{0}^{2\pi}r^{\sum_k\mft(k)-k}\frac{\Dt(\mft(\ul{K}))}{\Dt(\ul{K})}(e^{ix})^{-R+\sum_k\mft(k)-k}\,dx=0,
	\]
	unless $-R+\sum_k\mft(k)-k=0$. This gives a sum condition which all $\mft$ of the same size must satisfy. Likewise, 
	\[
	\rmn{Cr}_{\vec{y}}(\vec{g}_{\mft},x)=\det\left[y_k^{\mft(j)-1}\right]_{j,k=1}^K(e^{ix})^{-K+\sum_k\mft(k)}.
	\]
	Thus, 
	\[
	\int_{0}^{2\pi}\rmn{Cr}_{\vec{y}}(\vec{g}_{\mft},x)(e^{-ix})^{R}\,dx=\det\left[y_k^{\mft(j)-1}\right]_{j,k=1}^K\int_{0}^{2\pi}(e^{ix})^{-R-K+\sum_k\mft(k)}\,dx=0
	\]
	unless $-R-K+\sum_k\mft(k)=0$. This condition is actually quite strong and makes our $\gm(\vec{y})$ forms quite sparse. For example, when $K=2$, knowing $\mft(1)\in \ul{N}$ determines $\mft(2)=\mft(1)+R+2$, no matter how big $N$ is. 
	
	Historically, being able to ``diagonalize" the form $\gm$ by a clever choice of (potentially orthogonal or skew orthogonal) polynomials is incredibly useful in obtaining Pfaffian correlation functions from the Pfaffian partition functions. We expect this to be the case with Hyperpfaffian partition functions and correlation functions as well, though this is admittedly still speculation.

	\subsection{Limits of circular constellations}
	\label{ssec:limcirc}
	
	As in \autoref{sec:limlin}, we first consider limits (of Homogeneous Constellation Ensembles) as the distances between the circles shrinks to zero. The interaction terms which would give us singularities are the ones with $L_jL_k\log|y_ke^{ix_m}-y_je^{ix_m}|=L_jL_k\log(y_k-y_j)$, coming from particles which share an angle $x_m$. Thus, the correct denominator which accounts for these singularities is $\Dt^{\vec{L}}(\vec{y})$ so that 
	\[
	\lim_{\vec{y}\to \vec{1}}\frac{\Dt^{\textbf{L}}(\textbf{z})}{\left(\Dt^{\vec{L}}(\vec{y})\right)^M}=\Dt(e^{i\vec{x}})^{R_1^2},
	\]
	in which we take the limit as $y_1=\cdots=y_K=1$ to represent all the circles collapsing onto the unit circle. As before, the limiting Boltzmann factor corresponds to a one-dimensional ensemble of particles with charge $R_1=\sum_{k=1}^KL_k$. In terms of confluent matrices,
	\[
	\lim_{\vec{y}\to \vec{1}}\frac{V^{\textbf{L}}(\textbf{z})}{\left(\Dt^{\vec{L}}(\vec{y})\right)^M}=V^{R_1}(e^{i\vec{x}}),
	\]
	we get the same result as the linear case, with the location vector of real points replaced by a location vector of points on the unit circle. Additionally, in terms of the partition function, we get the same result as the linear case with $\rmn{Cr}(\vec{p}_t,x)$ in place of $\rmn{Wr}(\vec{p}_t,x)$. Explicitly,
	\[
	\lim_{\vec{y}\to \vec{1}}\frac{Z_M(\textbf{z})}{\left(\Dt^{\vec{L}}(\vec{y})\right)^M}=\frac{1}{M!}\int\left[\sum_{\mft:\ul{R_1}\nearrow\ul{N}}\int_{\R}\rmn{Cr}(\vec{p}_\mft,x)\,d\mu(x)\,\ep_{\mft}\right]^{\wedge M}\ep_{\rmn{vol}}
	\]
	when $R_1$ is even (and the analogous double-Wronskian expression holds when $R_1$ is odd). 
	
	Proceeding as we did in the linear case, we next consider limits as the distances between our circles increase without bound. For simplicity, we start by setting $y_k=1+hk$ (so that $\vec{y}\to \vec{1}$ as $h\to 0$) and then consider limits as $h\to \I$.  Recall (from \autoref{sec:circ}), there are three types of interaction terms in the Boltzmann factor. First, particles which share an angle $x_m$ produce an interaction factor of $(h(k-j))^{L_jL_k}$. These interactions are already accounted for by the $\left(\Dt^{\vec{L}}(\vec{y})\right)^M$ denominator. 
	
	Next, particles on the same circle of radius $y_k=1+hk$ produce an interaction factor of $\left((1+hk)\left(e^{ix_m}-e^{ix_n}\right)\right)^{L_k^2}$, which grows on the order of $(1+hk)^{L_k^2}$. There are ${M\choose 2}$ many of these for each $1\leq k\leq K$. Note, this is unique to the circular case, in which the particles drift apart as the radius of the circle grows without bound. Finally, particles at different angles on different circles produce an interaction factor of $\left((1+hk)e^{ix_m}-(1+hj)e^{ix_n}\right)^{L_jL_k}$, which grows on the order of $(h(k-j))^{L_jL_k}$. There are ${M\choose 2}{K\choose 2}$ many of these. Thus, if we set
	\[
	P_M^{\vec{L}}(h)=\left[\prod_{j\neq k}(1+h(k-j))^{L_jL_k}\prod_{k=1}^K(1+hk)^{L_k^2}\right]^{M\choose 2},
	\]
	then $\ds\lim_{h\to 0}P_M^{\vec{L}}(h)=1$, and it is straightforward to check
	\[
	\limi{h}\frac{\Dt^{\textbf{L}}(\textbf{z})}{\left(\Dt^{\vec{L}}(h\ul{K})\right)^MP_M^{\vec{L}}(h)}=\Dt(e^{i\vec{x}})^{L_1^2+\cdots+L_K^2}.
	\]
	However, the limitations of the linear case also apply in the circular case. In particular, the limiting partition function (as $h\to \I$) is still a limit of Hyperpfaffians rather than an honest Hyperpfaffian in its own right.

	\subsection{Multicomponent circular constellations}
	\label{ssec:mcirc}
	We should think of Multicomponent Circular Constellations as being variations on the linear case in which we plug in variables $\textbf{z}$ instead of $\textbf{x}$. At the beginning of \autoref{sec:circ}, we demonstrate we can account for the absolute value (complex modulus) by factoring out the sign corrections and then grouping them in place of the potential $U$. We do this in the previous volume as well for the one-dimensional (circular) ensembles. Alternatively, we can view Multicomponent Constellation Ensembles as being variations on the appropriate one-dimensional ensemble as in \autoref{ssec:constpf}. 
	
	Following the setup in \autoref{ssec:constpf}, let $\textbf{x}$ be the collection of angles $x_{m}^j\in [0,2\pi)$. Define $\textbf{x}_{\vec{y}}$ as before with all instances of $x_m^j+iy_k$ replaced with $y_ke^{ix_m^j}$. For particles $y_ke^{ix_m^j}$ and $y_ke^{ix_n^l}$ on the same circle, 
	\[
	\left|y_ke^{ix_m^j}-y_ke^{ix_n^l}\right|=-ie^{-i(x_m^j+x_n^l)/2}\left(y_ke^{ix_m^j}-y_ke^{ix_n^l}\right)\sgn(x_m^j-x_n^l).
	\]
	Using what we know from the one-dimensional case (Wolff and Wells 2021), the sign correction factors in $x_m^j$ are
	\[
	\left(-ie^{-ix_m^j}\right)^{KL_jT/2},
	\]
	where
	\[
	T=-L_j+\sum_{k=1}^{J}L_kM_k.
	\]
	Next, for particles $y_ke^{ix_m^j}$ and $y_le^{ix_m^j}$ which share an angle $x_m^j$,
	\[
	\left|y_ke^{ix_m^j}-y_le^{ix_m^j}\right|=e^{-ix_m^j}\left(y_ke^{ix_m^j}-y_le^{ix_m^j}\right),
	\]
	giving us the sign correction factor
	\[
	\left(e^{-ix_m^j}\right)^{L_j^2{K\choose 2}}.
	\]
	Finally, for particles $y_ke^{ix_m^j}$ and $y_he^{ix_n^l}$, which share neither an angle nor a radius,
	\[
	\left|y_ke^{ix_m^j}-y_he^{ix_n^l}\right|\left|y_he^{ix_m^j}-y_ke^{ix_n^l}\right|=-e^{-i(x_m^j+x_n^l)}\left(y_ke^{ix_m^j}-y_he^{ix_n^l}\right)\left(y_he^{ix_m^j}-y_ke^{ix_n^l}\right),
	\]
	giving us the last sign correction factor
	\[
	\left(-ie^{-ix_m^j}\right)^{{K\choose 2}L_jT}.
	\]
	Thus, 
	\[
	d\mu_j(x)=\left(-ie^{-ix}\right)^{K^2L_jT/2}\left(e^{-ix}\right)^{L_j^2{K\choose 2}}\,dx.
	\]
	Finally, we obtain the same Berezin integral expressions for the partition functions as the linear case (Lemma \ref{lem:canpf} for the Canonical, Theorems \ref{thm:isogcpf} and \ref{thm:allevens} for the Isocharge Grand Canonical) with new $\gm_{j}(\vec{y})$ and $\et_{j,k}(\vec{y})$ defined by
	\[
	\gm_j(\vec{y})=\sum_{\mft:\ul{L_jK}\nearrow \ul{N}}\int_{0}^{2\pi}\rmn{Wr}\otimes \rmn{Cr}_{\vec{y}}(\vec{p}_\mft,x)\,d\mu(x)\ep_{\mft},
	\]
	and
	\[\et_{j,k}(\vec{y})=\sum_{\mft:\ul{L_jK}\nearrow \ul{N}}\sum_{\mfs:\ul{L_kK}\nearrow \ul{N}}\int_{0}^{2\pi}\int_{0}^{2\pi}\rmn{Wr}\otimes \rmn{Cr}_{\vec{y}}(\vec{p}_\mft,x_1)\cdot \rmn{Wr}\otimes \rmn{Cr}_{\vec{y}}(\vec{p}_\mfs,x_2)\,d\mu(x_1)\,d\mu(x_2)\,\ep_{\mft}\wedge \ep_{\mfs}.
	\]

	\bibliography{icea}
	
\end{document}